\documentclass[a4paper,12pt]{article}
\pdfoutput=1
\usepackage[margin=1in]{geometry}
\usepackage{graphicx,xcolor}
\usepackage{mathtools,braket,blkarray}
\allowdisplaybreaks
\usepackage{amssymb,amsfonts,bbm,relsize}
\usepackage[width=0.95\textwidth,footnotesize]{caption}
\usepackage[font=scriptsize]{subcaption}
\usepackage{cite,hyperref,url}
\hypersetup{pageanchor=false}
\usepackage{booktabs,multirow,makecell}
\usepackage[utf8]{inputenc}
\usepackage{cleveref}
\usepackage{amsmath}

\begin{document}

\begin{titlepage}
\begin{center}
{\bf\Large Flipped $SU(5)$ with modular $A_4$ symmetry
  } \\[12mm]
Georgianna~Charalampous$^{a}$~\footnote{E-mail: \texttt{georgiana981@outlook.com}},
Stephen~F.~King$^{b}$~\footnote{E-mail: \texttt{king@soton.ac.uk}},
George K. Leontaris$^{a}$~\footnote{E-mail: \texttt{leonta@uoi.gr}},
Ye-Ling~Zhou$^{c,d}$~\footnote{E-mail: \texttt{zhouyeling@ucas.ac.cn}},
\\[-2mm]

\end{center}
\vspace*{0.50cm}
\centerline{$^{a}$~\it
			Physics Department, University of Ioannina, 45110, Ioannina, 	Greece}
 \vspace*{0.2cm}
\centerline{$^{b}$ \it
School of Physics and Astronomy, University of Southampton,}
\centerline{\it
SO17 1BJ Southampton, United Kingdom }
 \vspace*{0.2cm}
\centerline{$^{c}$ \it School of Fundamental Physics and Mathematical Sciences,} 
\centerline{\it Hangzhou Institute for Advanced Study, UCAS, Hangzhou, China}
 \vspace*{0.2cm}
\centerline{$^{d}$ \it International Centre for Theoretical Physics Asia-Pacific,} \centerline{\it Beijing/Hangzhou, China}
\vspace*{1.20cm}

\begin{abstract}
{\noindent
We study Flipped $SU(5)\times U(1)$ Grand Unified Theories (GUTs) with 
$\Gamma_3\simeq A_4$ modular symmetry. We propose two models with different modular weights assignments, where the fermion mass hierarchy can arise from weighton fields. In order to relax the constraint on the Dirac neutrino Yukawa matrix we 
appeal to mechanisms which allow incomplete GUT representations, allowing a good
fit to quark and charged lepton masses and quark mixing for a single modulus field
$\tau$, with the neutrino masses and lepton mixing well determined by the type I seesaw mechanism, at the expense of some tuning. We also discuss the double seesaw possibility allowed by the extra singlets generically predicted in such string inspired theories. 
}
\end{abstract}
\end{titlepage}

\section{Introduction}

 The existence  of  three  fermion families  and the origin of flavour mixing  are  long standing questions in particle physics.    Regarding the first issue,
there is no theoretical explanation  why there are three families for the otherwise successful standard model and its  field theory extensions such as Grand Unified Theories (GUT) and their supersymmetric analogues.  A possible interpretation gaining attention these days lies  in effective models derived within the context of string theory. 
In a wide class of such constructions the number of fermion generations is attributed to the topological properties -and in particular the Euler characteristic- of the compactification manifold. The origin of flavour mixing among the three families, however, is still unclear. Various mechanisms have been implemented, including those based on
 the geometry of the compactification manifold,   abelian and discrete symmetries, fluxes and non-perturbative effects, but they are still debatable.  

	Over the last few  decades  abelian and  non-abelian discrete symmetries have gained  an increasing interest  in  the particle physics literature, with special focus on  their r\^ole 
in model building. They have been introduced  to predict viable fermion
and particularly neutrino mass textures as well as to suppress processes leading to
baryon and lepton number violating interactions. 
 More recently, modular invariance has been suggested as a  novel candidate for predicting viable fermion mass textures~\cite{Feruglio:2017spp}~\footnote{For early work on mass matrices and modular  invariance in string motivated models see~\cite{Dudas:1995eq,Leontaris:1997vw,Dent:2001mn}}.
This is an intrinsic  symmetry in theories with ultra-violet  completion, such as string theory, and 
can exist in parallel with some discrete group of a different origin.
	Indeed, Modular Invariance is a fundamental concept  in string  theory and is naturally expected to leave its trace in the effective
field  theory model (possibly based on some GUT).  Among other implications, it governs the structure 
of the potential and particularly the Yukawa couplings. 
For example, in orientifold compactifications of Type II strings the Yukawa couplings are functions with specific modular  properties and in orbifold compactifications of heterotic strings  Yukawa couplings between twisted states  are subject to restrictions from modular invariance and recent attempts to exploit these properties  have already appeared~(see for example~\cite{Nilles:2020nnc,Kobayashi:2020uaj}).
In general, depending on the details of the derivation of the superstring model, Yukawa couplings are expected to be expressed in terms of certain modular forms  exhibiting certain transformation properties   under the modular group. 

Modular forms depend on a positive integer called the level $N$, together with an integer weight $k$, and are manifested as modular multiplets of the homogeneous finite modular group $\Gamma'_N\equiv\Gamma/\Gamma(N)$~\cite{Liu:2019khw}. If $k$ is an even number~\cite{Feruglio:2017spp}, they may be organised into modular multiplets of the inhomogeneous finite modular group $\Gamma_N\equiv\overline{\Gamma}/\overline{\Gamma}(N)$. 
Realistic models have been constructed based on $\Gamma_N$ 
for the levels
$N=2$~\cite{Kobayashi:2018vbk,Kobayashi:2018wkl,Kobayashi:2019rzp,Okada:2019xqk}, $N=3$~\cite{Feruglio:2017spp,Criado:2018thu,Kobayashi:2018vbk,Kobayashi:2018scp,deAnda:2018ecu,Okada:2018yrn,Kobayashi:2018wkl,Novichkov:2018yse,Nomura:2019jxj,Okada:2019uoy,Nomura:2019yft,Ding:2019zxk,Okada:2019mjf,Nomura:2019lnr,Kobayashi:2019xvz,Asaka:2019vev,Gui-JunDing:2019wap,Zhang:2019ngf,Nomura:2019xsb,Wang:2019xbo,Kobayashi:2019gtp,King:2020qaj,Ding:2020yen,Okada:2020rjb,Nomura:2020opk,Asaka:2020tmo,Okada:2020brs,Yao:2020qyy,King:2020qaj,Feruglio:2021dte}, $N=4$~\cite{Penedo:2018nmg,Novichkov:2018ovf,deMedeirosVarzielas:2019cyj,Kobayashi:2019mna,Criado:2019tzk,King:2019vhv}, \cite{Wang:2019ovr,Gui-JunDing:2019wap,Wang:2020dbp,Qu:2021jdy}, $N=5$~\cite{Novichkov:2018nkm, Ding:2019xna,Criado:2019tzk} and $N=7$~\cite{Ding:2020msi}. 
The modular invariance approach may also be extended to incorporate several factorizable~\cite{deMedeirosVarzielas:2019cyj} and non-factorizable moduli~\cite{Ding:2020zxw}.
Modular invariance can also address the origin of mass hierarchies without introducing an additional Froggatt-Nielsen (FN) $U(1)$~\cite{Froggatt:1978nt} symmetry.  The role of the FN flavon is played by a singlet field called the weighton~\cite{King:2020qaj}, which carries a non-zero modular weight, but no other charges. 

Grand unified theories (GUTs)
are well motivated theories which reduce the three gauge interactions of the SM group
into a simpler structure such as $SU(5)$~\cite{Georgi:1974sy}. In GUTs,
the quark and lepton fields are embedded into fewer gauge multiplets, resulting in relations between 
quark and lepton mass matrices. There is good motivation for including also 
family symmetry together with GUTs in order to account for large lepton mixing~\cite{King:2017guk}. 
The discrete group $A_4$ is the minimal choice which admits triplet representations~\cite{Ma:2001dn}.
Without modular symmetry, combining $A_4$ family symmetry with $SU(5)$ GUTs~\cite{Bjorkeroth:2015ora}
requires vacuum alignment of the flavons in order to break the $A_4$, and this provides 
a further motivation for including modular symmetry. Modular symmetry was first 
combined with $SU(5)$ GUTs in an $(\Gamma_3\simeq A_4)\times SU(5)$ model
in~\cite{deAnda:2018ecu,Chen:2021zty}, and subsequently modular $SU(5)$ GUT models have been constructed based on
$(\Gamma_2\simeq S_3)\times SU(5)$~\cite{Kobayashi:2019rzp,Du:2020ylx}, and $(\Gamma_4\simeq S_4)\times SU(5)$~\cite{King:2021fhl,Zhao:2021jxg,Ding:2021zbg}.
Recently $SO(10)$ GUTs have been studied based on $\Gamma_3\simeq A_4$ modular symmetry
\cite{Ding:2021eva}.

The above studied GUTs depend on a Higgs in an adjoint representation in order to break the gauge symmetry,
unless an extra-dimensional mechanism is invoked such as Wilson lines or F-theory flux breaking.
On the other hand, in some string theories the adjoint representation is not available
to break GUT symmetry, for example the viable GUT models in the context of heterotic 
compactifications, are only those which do not rely on such Higgs fields.
The most popular ones  are the  Flipped $SU(5)$ \cite{DeRujula:1980qc,Georgi:1980pw,Barr:1981qv,Derendinger:1983aj,Antoniadis:1987dx} and the Pati-Salam models \cite{Pati:1974yy,Antoniadis:1988cm}

In this paper, motivated by the above considerations, we study Flipped $SU(5)\times U(1)$ GUTs with 
$\Gamma_3\simeq A_4$ modular symmetry. To illustrate the approach, we propose two models with different modular weights assignments, where the fermion mass hierarchy can arise from weighton fields, where one of the models is studied in detail using a 
numerical $\chi^2$ analysis. In such models the neutrino sector can be tightly constrained by the up type quark mass matrix, in particular the up-quarks and Dirac neutrino mass matrices satisfy the relation $m_D^T=m_u$ at the GUT scale.
In order to avoid this constraint we appeal to F-theory constructions where the components of the GUT multiplets may 
lie on different matter curves. With this constraint relaxed, we can fit the quark and lepton mass matrices and quark mixing for a single modulus field
$\tau$, with the neutrino masses and lepton mixing determined by the type I seesaw mechanism. We also discuss the double seesaw possibility allowed by the extra singlets possible in Flipped $SU(5)$. 

The layout of the remainder of the paper is as follows. In section 2 we 
present a short introduction to modular transformations, mainly  focusing on the modular symmetry $A_4$. In section 3 we start a brief account of the
field theory version of the Flipped $SU(5)$ model. Next, we proceed with
 a modular invariant  version proposed in the present work. Considering that 
Yukawa couplings are certain modular forms, we derive the modular invariant
superpotential and the induced mass matrices for up and down quarks, charged  leptons and neutrinos. We perform an detailed numerical investigation and 
show that the proposed construction is in agreement with all low energy data regarding the fermion masses and their mixing. In section 4 we discuss the predictions of the suggested models and summarise the main results.  We study the contribution of singlet neutrinos to the flavour structure in Appendix A. A variant of this model, based on a different choice of the modular properties is presented in the Appendix B.


\section{Modular symmetries}

 
In this section, we give a brief review of the modular symmetry and the tetrahedral group $A_4$ as a finite modular group. 

\subsection{The infinite modular symmetry}

A modular transformation $\gamma$ is defined as a linear fractional transformation on the complex modulus $\tau$ varying in the upper-half plane ${\cal H} = {\rm Im}(\tau) > 0$,
\begin{equation}
\gamma:\quad
\tau \to \gamma \tau =\frac{a\tau +b}{c\tau +d} \,,
\end{equation}
where $a, b, c, d$ are integers and $ad-bc=1$.  Each modular transformation can be represented by a $2\times 2$ matrix with integer  entries and the determinant equal to one, i.e.,
\begin{eqnarray}
\gamma = \begin{pmatrix}
a & b \\
c & d \\
\end{pmatrix}\,, \quad
\det(\gamma) =1 \,.
\end{eqnarray} 
The modular group $\bar\Gamma$ is defined as a group of these transformations, i.e.,
\begin{eqnarray}
\bar{\Gamma} = \left\{ \begin{pmatrix}
a & b \\
c & d \\
\end{pmatrix} /(\pm \mathbf{1})
\Big|a,b,c,d \in \mathbb{Z}, ad-bc=1 \right\}  \,.
\end{eqnarray}
It includes infinite elements. All elements can be generated by $S$ and $T$, given by
\begin{eqnarray}
S:\tau \longmapsto -\dfrac{1}{\tau} \,,
\quad
T:\tau \longmapsto \tau+1 \,.
\end{eqnarray}
They are represented by $2\times 2$ matrices as
\begin{eqnarray}
S=\begin{pmatrix}
0 & 1 \\
-1 & 0 \\
\end{pmatrix} \,,
\quad
T=
\begin{pmatrix}
1 & 1 \\
0 & 1 \\
\end{pmatrix} \,.
\end{eqnarray}
The actions of $S$ and $T$ in ${\cal H}$ are given by:
\begin{eqnarray}
S:\tau \longmapsto -\dfrac{1}{\tau} \,,
\quad
T:\tau \longmapsto \tau+1 \,.
\end{eqnarray}
One can prove that the identities $S^{2}=(ST)^{3}=\mathbb{I}$ are satisfied, namely, $S^{2}\tau=(ST)^{3} \tau=\tau$. 

A subgroup of $\overline{\Gamma}$ is obtained by restricting $a, d = 1~({\rm mod}~N)$ and $b, c =  0~({\rm mod}~N)$,  
\begin{eqnarray}
\overline{\Gamma}(N) = \left\{ \begin{pmatrix} a & b \\ c & d \end{pmatrix} \in \bar \Gamma, ~~  \begin{pmatrix} a & b \\ c & d \end{pmatrix} = \begin{pmatrix} 1 & 0 \\ 0 & 1 \end{pmatrix} ~~ ({\rm mod}~ N) \right\} \,,
\end{eqnarray}
where $N$ is a positive integer.
$\overline{\Gamma}(N)$ is also an infinite group. 
The quotient group $\overline{\Gamma}/\overline{\Gamma}(N)$ is finite and labelled as $\Gamma_N$. It is equivalently obtained by requiring $a,b,c,d \in \mathbb{Z}_N$, namely  
\begin{eqnarray}
\Gamma_N= \left\{ \begin{pmatrix}
a & b \\
c & d \\
\end{pmatrix}/(\pm \mathbf{1})
\Big|a,b,c,d \in \mathbb{Z}_N, ad-bc=1 \right\} \,.
\end{eqnarray}
For $N=2,3,4,5,7$, $\Gamma_2 \simeq S_3$, $\Gamma_3 \simeq A_4$, $\Gamma_4 \simeq S_4$, $\Gamma_5 \simeq A_5$ and $\Gamma_7 \simeq \Sigma(168)$. 

In ${\cal N} =1$ supersymmetric theories, a modular-invariant superpertential is expanded as series of polynomials in powers of supermultiplets $\phi^{I}$,
\begin{eqnarray}
{\cal W} = \sum_n Y_{I_1 I_2 \cdots I_n}(\tau) \phi^{I_1} \phi^{I_2} \cdots \phi^{I_n} \,.
\end{eqnarray}
Here, $Y_{I_1 I_2 \cdots I_n}(\tau)$ are called modular forms keeping the superpotential term invariant under any modular transformation. 
A modular form $Y_i(\tau)$ of level $N$ and weight $2k$ is defined as a holomorphic function of the modulus $\tau$ with the modular transformations under $\Gamma(N)$,
\begin{equation}
\gamma \in \Gamma(N): \quad Y_i(\tau) \to Y_i(\gamma \tau)= (c\tau + d)^{2k} \hspace{0.06cm} Y_i(\tau).  
\end{equation}
Under the quotient group $\Gamma_N$, they are transformed not as holomorphic functions but linear superposition of a series of  modular forms $\{  Y_1(\tau), Y_2(\tau),\cdots\}$ which take the same weight and level, 
\begin{equation}
\gamma \in \Gamma_N: \quad Y_i(\tau) \to Y_i(\gamma \tau)= (c\tau + d)^{2k} \sum_j \rho_{ij}(\gamma) \hspace{0.06cm} Y_j (\tau), 
\end{equation} 
where $j$ runs as an index of the series $\{  Y_1(\tau), Y_2(\tau),\cdots\}$, $\rho(\gamma)$ is a unitary representation matrix of $\gamma \in \Gamma_N$. For a given finite $\Gamma_N$, one can choose a basis, where the representation $\rho$ is decomposed to a few irreducible representations. In this basis, modular forms and fields appear as a series of irreducible representions of $\Gamma_N$. Assigning this basis as the flavour basis of matter fields, the restriction of the modular invariance  could strongly constrain the flavour structure. In this work, we will take $\Gamma_3 \simeq A_4$ as an example to discuss the flavour mixing in Flipped $SU(5)$ framework. 

\subsection{The finite modular symmetry $A_4$}

$\Gamma_3$ is a finite subgroup of $\bar \Gamma$, referring to $N=3$. 
Due to the requirement $a,b,c,d \in \mathbb{Z}_3$, the generator $T$ satisfies one more condition $T^{3}=\mathbb{I}$, leading to its isomorphism to the tetrahedral group $A_4$. 

$A_4$ contains 12 elements. All can be written as products of $S$ and $T$ and are shown below,
\begin{eqnarray}
 I, T, ST, TS, STS, T^2, ST^2, T^2S, TST, S, T^2ST, TST^2 \,.
\end{eqnarray} 
It has three singlet ($\mathbf{1}$, $\mathbf{1}'$ and $\mathbf{1}''$) and one triplet ($\mathbf{3}$) irreducible representations. The generators $S$ and $T$ in these representations are given by
\begin{eqnarray} \label{eq:basis}
\mathbf{1}\;: \quad&& S=1\,,\hspace{3cm} T = 1 \,, \nonumber\\
\mathbf{1}'\,: \quad&& S=1\,,\hspace{3cm} T = \omega \,, \nonumber\\
\mathbf{1}'': \quad&& S=1\,,\hspace{3cm} T = \omega^{2} \,, \nonumber\\
\mathbf{3}\;: \quad&& S=\dfrac{1}{3}
\begin{pmatrix}
-1 & 2 & 2 \\
2 & -1 & 2 \\
2 & 2 & -1 \\
\end{pmatrix} \,, \quad
T=
\begin{pmatrix}
1 & 0 & 0 \\
0 & \omega & 0 \\
0 & 0 & \omega^{2} \\
\end{pmatrix} \,,
\end{eqnarray}
where $\omega=e^{2i \pi/3}$. Tensor products of two  irreducible representations are decomposed as,
\begin{eqnarray}
&&\mathbf{1} \otimes \mathbf{r}=\mathbf{r}, \hspace{0.5cm} 
\mathbf{1'} \otimes \mathbf{1'}= \mathbf{1''}, \hspace{0.5cm} 
\mathbf{1''} \otimes \mathbf{1''}=\mathbf{1'}, \hspace{0.5cm} 
\mathbf{1'} \otimes \mathbf{1''} = \mathbf{1}, \nonumber\\
&&\mathbf{1'} \otimes \mathbf{3}= \mathbf{3}, \hspace{0.5cm} 
\mathbf{1''} \otimes \mathbf{3}=\mathbf{3}, \hspace{0.5cm} 
\mathbf{3} \otimes \mathbf{3}' =  \mathbf{1} \oplus \mathbf{1}' \oplus \mathbf{1}'' \oplus \mathbf{3}_S \oplus \mathbf{3}_A \,,
\end{eqnarray}
where $\mathbf{r} = \mathbf{1}, \mathbf{1}', \mathbf{1}'', \mathbf{3}$, and ``$S$'' and ``$A$'' in the subscript denote symmetric and antisymmetric combinations, respectively.
In particular, the decomposition of the tensor product of two triplets $a=(a_1, a_2, a_3)^T$ and $b = (b_1, b_2, b_3)^T$ is explicitly written as:
\begin{eqnarray}
(ab)_{\mathbf{1}} &=& a_{1}b_{1}+a_{2}b_{3}+a_{3}b_{2} \,,
\nonumber\\
(ab)_{\mathbf{1'}} &=& a_{3}b_{3} + a_{1}b_{2} +a_{2}b_{1} \,,
\nonumber
\\
(ab)_{\mathbf{1''}} &=& a_{2}b_{2}+a_{1}b_{3}+a_{3}b_{1} \,,
\nonumber
\\
(ab)_{\mathbf{3}_{S}} &=& \begin{pmatrix}
2a_{1}b_{1}-a_{2}b_{3}-a_{3}b_{2} \\
2a_{3}b_{3}-a_{1}b_{2}-a_{2}b_{1} \\
2a_{2}b_{2}-a_{1}b_{3}-a_{3}b_{1} \\
\end{pmatrix}
 \,,
\nonumber\\
(ab)_{\mathbf{3}_{A}} &=& 
\begin{pmatrix}
a_{2}b_{3}-a_{3}b_{2} \\
a_{1}b_{2}-a_{2}b_{1} \\
a_{3}b_{1}-a_{1}b_{3} \\
\end{pmatrix} \,.
\end{eqnarray}

Modular forms of level $N=3$ and weight $2k$ form a linear space of dimension $2k+1$. All of them have been explicitly obtained in
terms of the Dedekind eta-function $\eta(\tau)$ \cite{Feruglio:2017spp}:
\begin{equation}
\eta(\tau)=q^{1/24} \prod_{n=1}^{\infty} (1-q^{n}), \hspace{0.4cm} q=e^{2 \pi i \tau}
\end{equation}
\begin{itemize}
	\item
For $2k=2$, 
there are  3 linearly independent modular forms, transforming as a triplet of $A_4$. Given the triplet basis in Eq.~\eqref{eq:basis}, this triplet modular form is written as $Y_{\textbf{3}}^{(2)}=(Y_{1},Y_{2},Y_{3})^{T} \sim \mathbf{3}$ with
 \begin{eqnarray}
Y_{1}(\tau) &=& \dfrac{i}{2 \pi} \left[\dfrac{\eta'(\tau/3)}{\eta(\tau/3)}+\dfrac{\eta'((\tau+1)/3)}{\eta((\tau+1)/3)} + \dfrac{\eta'((\tau+2)/3)}{\eta((\tau+2)/3)} - \dfrac{27 \eta'(3\tau)}{\eta(3\tau)}\right] \,, 
\nonumber\\
Y_{2}(\tau) &=& \dfrac{-i}{\pi} \left[\dfrac{\eta'(\tau/3)}{\eta(\tau/3)}+\omega^{2} \dfrac{\eta'((\tau+1)/3)}{\eta((\tau+1)/3)} + \omega \dfrac{\eta'((\tau+2)/3)}{\eta((\tau+2)/3)}\right] \,,
\nonumber\\
Y_{3}(\tau) &=& \dfrac{-i}{\pi} \left[\dfrac{\eta'(\tau/3)}{\eta(\tau/3)}+\omega \dfrac{\eta'((\tau+1)/3)}{\eta((\tau+1)/3)} + \omega^{2} \dfrac{\eta'((\tau+2)/3)}{\eta((\tau+2)/3)}\right] \,.
\end{eqnarray}
Modular forms of higher weights are derived from products of those of lower weights.  
\item
For $2k=4$,  there are 5 linearly independent modular forms, derived from products of two modular forms of weight $2$ and written as 
\begin{eqnarray}
Y_{\textbf{3}}^{(4)} &=& (Y_{\textbf{3}}^{(2)}Y_{\textbf{3}}^{(2)})_{\textbf{3}}=
\begin{pmatrix}
Y_{1}^{(4)} \\
Y_{2}^{(4)} \\
Y_{3}^{(4)} \\
\end{pmatrix}
=
\begin{pmatrix}
Y_{1}^{2}-Y_{2}Y_{3}
\\
Y_{3}^{2}-Y_{1}Y_{2}
\\
Y_{2}^{2}-Y_{1}Y_{3}
\end{pmatrix} \,,
\nonumber\\
Y_{\textbf{1}}^{(4)} &=& (Y_{\textbf{3}}^{(2)}Y_{\textbf{3}}^{(2)})_{\textbf{1}}=Y_{1}^{2} + 2Y_{2}Y_{3} \,,
\nonumber\\
Y_{\textbf{1$'$}}^{(4)} &=& (Y_{\textbf{3}}^{(2)}Y_{\textbf{3}}^{(2)})_{\textbf{1$'$}}=Y_{3}^{2} + 2Y_{1} Y_{2} \,.
\end{eqnarray} 
\item
For $2k=6$, the 7 linearly independent modular forms are given by 
\begin{eqnarray}
Y_{\textbf{1}}^{(6)} &=& (Y_{\textbf{3}}^{(2)}Y_{\textbf{3}}^{(4)})_{\textbf{1}}=Y_{1}^{3} + Y_{2}^{3} +Y_{3}^{3}-3Y_{1}Y_{2}Y_{3} \,,
\nonumber\\
Y^{(6)}_{\textbf{3}_{1}} &=& Y_{\textbf{3}}^{(2)}Y_{\textbf{1}}^{(4)}=
\begin{pmatrix}
Y_{1_{1}}^{(6)} \\
Y_{2_{1}}^{(6)} \\
Y_{3_{1}}^{(6)} \\
\end{pmatrix}
=
\begin{pmatrix}
Y_{1}^{3}+2Y_{1}Y_{2}Y_{3} \\
Y_{1}^{2}Y_{2}+2Y_{2}^{2}Y_{3} \\
Y_{1}^{2}Y_{3}+2Y_{3}^{2}Y_{2} \\
\end{pmatrix} \,,
\nonumber\\
Y_{\textbf{3}_{2}}^{(6)} &=& Y_{\textbf{3}}^{(2)}Y_{\textbf{1$'$}}^{(4)} =
\begin{pmatrix}
Y_{1_{2}}^{(6)} \\
Y_{2_{2}}^{(6)} \\
Y_{3_{2}}^{(6)} \\
\end{pmatrix}
=
\begin{pmatrix}
Y_{3}^{3}+2Y_{1}Y_{2}Y_{3} \\
Y_{3}^{2}Y_{1}+2Y_{1}^{2}Y_{2} \\
Y_{3}^{2}Y_{2}+2Y_{2}^{2}Y_{1} \\
\end{pmatrix} \,.
\end{eqnarray}
\item
For $2k=8$, the 9 linearly independent modular forms are  
\begin{eqnarray}
Y_{\mathbf{1}}^{(8)} &=& (Y_{\mathbf{3}}^{(2)}Y_{\mathbf{3}_{1}}^{(6)})_{\mathbf{1}}= (Y_{1}^{2}+2Y_{2}Y_{3})^{2}  \,,
\nonumber\\
Y_{\mathbf{1'}}^{(8)} &=& (Y_{\mathbf{3}}^{(2)}Y_{\mathbf{3}_{1}}^{(6)})_{\mathbf{1'}}=(Y_{1}^{2}+2Y_{2}Y_{3})(Y_{3}^{2}+2Y_{1}Y_{2}) \,,
\nonumber\\
Y_{\mathbf{1''}}^{(8)} &=& (Y_{\mathbf{3}}^{(2)}Y_{\mathbf{3}_{2}}^{(6)})_{\mathbf{1''}}=(Y_{3}^{2}+2Y_{1}Y_{2})^{2} \,,
\nonumber\\
Y_{\mathbf{3}_{1}}^{(8)} &=& Y_{\mathbf{3}}^{(2)}Y_{\mathbf{1}}^{(6)}=\begin{pmatrix}
Y_{1_{1}}^{(8)} \\
Y_{2_{1}}^{(8)} \\
Y_{3_{1}}^{(8)} \\
\end{pmatrix}
= (Y_{1}^{3}+Y_{2}^{3}+Y_{3}^{3}-3Y_{1}Y_{2}Y_{3})
\begin{pmatrix}
Y_{1} \\
Y_{2} \\
Y_{3} \\
\end{pmatrix} \,,
\nonumber\\
Y_{\mathbf{3}_{2}}^{(8)} &=& (Y_{\mathbf{3}}^{(2)}Y_{\mathbf{3}_{2}}^{(6)})_{\mathbf{3}_{A}}=\begin{pmatrix}
Y_{1,_{2}}^{(8)} \\
Y_{2_{2}}^{(8)} \\
Y_{3_{2}}^{(8)} \\
\end{pmatrix}
= (Y_{3}^{2}+2Y_{1}Y_{2})
\begin{pmatrix}
(Y_{2}^{2}-Y_{1}Y_{3}) \\
(Y_{1}^{2}-Y_{2}Y_{3}) \\
(Y_{3}^{2}-Y_{1}Y_{2}) \\
\end{pmatrix} \,.
\end{eqnarray}
\end{itemize}
We further list some singlet modular forms of higher weights ( $2k=10$ and $12$), which may be useful for the rest of the work, 
\begin{eqnarray}
Y_{\bf 1}^{(10)} &=& (Y_{\mathbf{3}}^{(2)}Y_{\mathbf{3}_{1}}^{(8)})_{1}=(Y_{1}^{3}+Y_{2}^{3}+Y_{3}^{3}-3Y_{1}Y_{2}Y_{3})(Y_{1}^{2}+2Y_{2}Y_{3}) \,,
\nonumber\\
Y_{\bf 1'}^{(10)} &=& (Y_{\mathbf{3}}^{(2)}Y_{\mathbf{3}_{1}}^{(8)})_{1'}=(Y_{1}^{3}+Y_{2}^{3}+Y_{3}^{3}-3Y_{1}Y_{2}Y_{3})(Y_{3}^{2}+2Y_{1}Y_{2}) \,,
\nonumber\\
Y_{\bf 1''}^{(10)} &=& (Y_{\mathbf{3}}^{(2)} Y_{\mathbf{3}_{2}}^{(8)})_{1''}= (Y_{3}^{2}+2Y_{1}Y_{2})[Y_{3}(Y_{2}^{2}-Y_{1}Y_{\bf 3})+Y_{2}(Y_{1}^{2}-Y_{2}Y_{3})+Y_{1}(Y_{3}^{2}-Y_{1}Y_{2})] \,, \nonumber\\
Y_{\mathbf{1}}^{(12)} &=& (Y_{\mathbf{3}}^{(4)}Y_{\mathbf{3}_{1}}^{(8)})_{\mathbf{1}}=(Y_{1}^{3}+Y_{2}^{3}+Y_{3}^{3}-3Y_{1}Y_{2}Y_{3})^2 \,,
\nonumber\\
Y_{\mathbf{1}}^{(12)} &=& (Y_{\mathbf{1}}^{4})^{3}=(Y_{1}^{2}+2Y_{2}Y_{3})^{3} \,,
\nonumber \\
Y_{\mathbf{1'}}^{(12)} &=& (Y_{\mathbf{1}}^{(4)})^{2}Y_{\mathbf{1'}}^{(4)}=(Y_{1}^{2}+2Y_{2}Y_{3})^{2}(Y_{3}^{2}+2Y_{1}Y_{2}) \,,
\nonumber\\
Y_{\mathbf{1''}}^{(12)} &=& (Y_{\mathbf{3}}^{(4)}Y_{\mathbf{3}_{2}}^{(8)})_{\mathbf{1''}}=(Y_{3}^{2}+2Y_{1}Y_{2})[2(Y_{3}^{2}-Y_{1}Y_{2})(Y_{1}^{2}-Y_{2}Y_{3})+(Y_{2}^{2}-Y_{1}Y_{3})^{2}] \,.
\end{eqnarray}

\section{Model Building}


\subsection{The Flipped $SU(5)$ framework}

The Flipped  $SU(5)$ model has been proposed  long time ago~\cite{Barr:1981qv, Derendinger:1983aj} as an alternative symmetry breaking pattern of the $SO(10)$ gauge group.
It is based on the $SU(5)\times U(1)_{\chi}$  gauge symmetry and has been reconsidered as a possible superstring
alternative to Georgi-Glashow $SU(5)$ due to  the fact that its spontaneous breaking to SM symmetry requires only a pair of ${\bf 10}+\overline{\bf 10}$ Higgs representations and does not need any adjoint Higgs representation. 
 In fact, this is a welcome property since in many string derived 
 effective models  the Higgs adjoint representation does not appear in the massless spectrum. 
 Among other virtues the model admits a doublet-triplet mass splitting for
 the color triplets, and in the presence of additional neutral singlets,  an extended  seesaw mechanism for neutrino masses is naturally realised. 
The hypercharge  generator is a linear combination of the $U(1)$ inside $SU(5)$ and the external abelian factor $ U(1)_{\chi}$  and  it is 
 no longer fully embedded in $SU(5)$. This way 
Flipped $SU(5)$  representations accommodate the SM matter fields  differently.  To start with,  the following quantum numbers follow from  $SO(10)\to SU(5)\times U(1)_{\chi}$ decompositions 
\begin{eqnarray}  
	\mathbf{16}&\to& (\mathbf{10},-\frac 12) +(\bar{\mathbf{5}},{\frac 32})+(\mathbf{1},{-\frac 52}) \,,
	\nonumber\\
	\mathbf{10}&\to& (\mathbf{5},1)+(\bar{\mathbf{5}},{-1}) \,.
\end{eqnarray} 

The Flipped  gauge symmetry $SU(5) \times U(1)_\chi$,  can be broken to the SM gauge symmetry via a two-step symmetry breaking,  $SU(5) \times U(1)_\chi \to SU(3)_c\times SU(2)_L \times U(1)_y \times U(1)_\chi \to  SU(3)_c\times SU(2)_L \times U(1)_Y$. In the first step of symmetry breaking, representation decompositions follow as, 
\begin{eqnarray}  
	(\mathbf{10},{-\frac 12}) &\to& (\mathbf{3},\mathbf{2},{\frac 16, -\frac 12}) +(\bar{\mathbf{3}},\mathbf{1},{-\frac 23, -\frac 12})+(\mathbf{1},\mathbf{1},{1, -\frac 12}) \,,
	\nonumber\\
	(\bar{\mathbf{5}},{\frac 32}) &\to&	(\bar{\mathbf{3}}, \mathbf{1}, \frac 13, \frac 32) + (\mathbf{1}, \mathbf{2}, -\frac 12, \frac 32) \,,
	\nonumber\\
	(\bar{\mathbf{5}},{-1}) &\to&	(\bar{\mathbf{3}}, \mathbf{1}, \frac 13, -1) + (\mathbf{1}, \mathbf{2}, -\frac 12, -1) \,.
\end{eqnarray} 
In the second step of symmetry breaking, the two $U(1)$ symmetries are broken to $U(1)_Y$ with 
the hypercharge defined by
\begin{eqnarray} 
Y=-\frac 15\left( y+ 2\chi \right) \,.
\end{eqnarray} 
Each representations then gain hypercharges as
\begin{eqnarray}  
(\mathbf{10},{-\frac 12}) &\to& Y=\{\frac 16,\frac 13, 0\} \to \{Q,d^c,\nu^c\} \,,
\nonumber\\
(\bar{\mathbf{5}},{+\frac 32}) &\to& Y=\{-\frac 23, -\frac 12\}\to \{u^c,L\} \,,
\nonumber\\
(\mathbf{1},{-\frac 52}) &\to& Y=\{+1\}\to e^c \,,
\nonumber\\
(\bar{\mathbf{5}},{-1}) &\to& Y=\{\frac 13, \frac 12\}\to \{D^c,h_u\} \,,
\nonumber\\
(\mathbf{5},{+1}) &\to& Y=\{-\frac 13, -\frac 12\}\to \{D, h_d\} \,,
\end{eqnarray} 
where $Q=(u,d)$ and $L = (\nu, e)$. 
After the symmetry breaking, SM matter fields $Q, L, u^c, d^c, e^c$, as well as the right-handed neutrino $\nu^c$ and MSSM Higgses $h_u$ and $h_d$, are seperated as shown on the right-hand side of the above formula. 
The definition of the hypercharge includes a component of the
external $U(1)_{\chi}$ in such a way that flips the positions of $u^c\leftrightarrow  d^c$ and $e^c\leftrightarrow  \nu^c$ within these representations,
while leaves the remaining unaltered. 

In summary, we obtain the following `Flipped' embedding of the SM representations. The chiral matter fields are 
\begin{eqnarray}
F_i&=&(\mathbf{10},{-\frac 12})\;=\;\{Q_i,d^c_i,\nu^c_i\} \,,\nonumber\\
\bar f_i&=&(\bar{\mathbf{5}},{+\frac 32})\;=\;\{u^c_i,L_i\} \,,\nonumber\\
\ell^c_i&=&(\mathbf{1},{-\frac 52})\;=\;e^c_i \,.
\end{eqnarray}
The Higgs fields breaking GUT and SM symmetries  reside  in the following 
Flipped $SU(5)$ representations 
\begin{eqnarray}
H\equiv (\mathbf{10},{-\frac 12})\;=\;\{Q_H,D_H^c,\nu_H^c\}&,&\bar H\equiv (\overline{\mathbf{10}},{+\frac 12})\;=\;\{\bar Q_H,\bar d_H^c,\bar \nu_H^c\} \,, \nonumber\\
h\equiv (\mathbf{5},{+1})\;=\;\{D_h,h_d\}&,&\bar h\equiv (\bar{\mathbf{5}},{-1})\;=\;\{\bar D_h,h_u\}\,.
\end{eqnarray}
A remarkable fact in the Flipped model, is  that
the $\bar{\mathbf{5}}$ matter field is completely distinguished from the $\bar{\mathbf{5}}$ Higgs field.
Indeed, due to their different $U(1)_{\chi }$ charge which is involved in the
hypercharge definition,  their  SM components
do not contain exactly the same type of SM-fields (the $\bar{\mathbf{5}}$ matter field contains $u^c$, while the  $\bar{\mathbf{5}}$ Higgs field contains the down-type $D_h$).
 Several $R$-parity violating
terms  will not be allowed because of this distinction.

The fermion masses arise from the following $SU(5)\times U(1)_{\chi}$ invariant couplings
\begin{eqnarray}
\label{eq:FermCoupl}
{\cal W}_d&=&(\mathbf{10},{-\frac 12 }) \cdot (\mathbf{10},{-\frac 12}) \cdot (\mathbf{5},{1})\;\to \; Q\,d^c\,h_d \,, \nonumber\\
{\cal W}_u&=&(\mathbf{10},{-\frac 12}) \cdot (\bar{\mathbf{5}},{\frac 32}) \cdot (\bar{\mathbf{5}},{-1})\;\to \;Q\,u^c\,h_u+\nu^c\,L\, h_u \,, \nonumber\\
{\cal W}_l&=&(\mathbf{1},{-\frac 52}) \cdot (\bar{\mathbf{5}}, {\frac 32}) \cdot (\mathbf{5},1)\;\rightarrow \; e^c\,L \,h_d \,.
\end{eqnarray}
Also, a higher order term providing Majorana masses  for the right-handed neutrinos can be written
\begin{eqnarray}
{\cal W}_{\nu^c}&=&\lambda^{\nu^c}_{ij}\frac{1}{M_S}\,{\bar H}\,{\bar H}\,{F_i}\,F_j\to \lambda^{\nu^c}_{ij}\frac{\langle \bar\nu^c_H\rangle^2}{M_S}\nu^c_i\nu^c_j \,.
\label{Maj}
\end{eqnarray}
If additional singlet fields $\nu_S, \Phi_i$ are present (which is the usual case in 
string derived models), then -depending on their specific properties- the
following couplings could be generated
\begin{eqnarray}  
F\bar H \nu_S+ \bar h h \Phi_i + \lambda_{ijk} \Phi_i\Phi_j\Phi_k
+\cdots 
\end{eqnarray}
In addition to SM representations, the Higgs sector contains 
dangerous colour triplets $D_h, D_H+c.c.$. They become massive 
through the following terms
\begin{eqnarray}  
HHh +\bar H\bar H\bar h \to  \langle \nu^c_H\rangle  D_H^c D_h
+ \langle \bar \nu^c_H\rangle \bar D_H^c  \bar  D_h \,.
\end{eqnarray} 

Note that, if no other symmetry exists the  terms such as 
$H F_i h,  H \bar f_j \bar h$ could be possible. Such terms would generate
dangerous mixing bewteen Higgs and Matter:
\begin{eqnarray}    (a F+b  H) \bar f_j \bar h +\cdots
\end{eqnarray} 
Remarkably, a $Z_2$ symmetry~\cite{Antoniadis:1987dx} which is odd only for $H\to -H$ 
excludes all these couplings from the lagrangian, while all the previous (useful)
terms are left intact. Similar symmetries have been discussed in~\cite{Kyae:2005nv}

For rank one mass textures the couplings in Eq.~\eqref{eq:FermCoupl} predict $m_t=m_{\nu_{\tau}}$ at the GUT scale.
However, in contrast to the standard $SU(5)$ model, down quark and lepton mass matrices are not
related, since at the $SU(5)\times U(1)_{\chi}$ level they originate from different Yukawa
couplings. This is an important difference with the ordinary $SU(5)$. We know that
in order to obtain the observed lepton and down quark mass spectrum at low energies,
at the GUT scale the following relations should hold \cite{Georgi:1979df}
\begin{eqnarray}
m_{\tau}=m_b\,,\quad
m_{\mu}=3\,m_s\,.
\end{eqnarray}
In the ordinary $SU(5)$, the masses are related and the relations
can be attributed to the Higgs adjoint which couples differently.  This mechanism though
is not operative in Flipped $SU(5)$ due to the absence of the adjoint, as noticed above, thus 
the mass matrices are not related and Yukawas can be adjusted accordingly.

We further review the derivation of the matching condition between gauge couplings of $U(1)_Y$ and those of $SU(5) \times U(1)_\chi$. Computing the traces and finding normalisation constants so that 
final trace is 2 give
\begin{eqnarray} 
C_y^2y^2&=&C_y^2 \frac{10}3 =2 \to C_y=\sqrt{\frac 35} \,,
\nonumber\\
C_{\chi}^2\chi^2&=&C_\chi^2 20 =2 \to C_{\chi}=\frac{1}{\sqrt{10}} \,.
\end{eqnarray} 
In terms of normalised  generators, 
$\tilde Y= \frac{1}{5C_y}\left( \tilde y+\kappa \tilde\chi\right)$
where the ratio is 
$\kappa\equiv 2 \frac{C_{y}}{C_{\chi}} = 2\sqrt{6} $. 
Finally  $Y=\sqrt{\frac 35}\tilde Y$ implies
\begin{eqnarray} 
Y= \frac 15 \left( \tilde y+2 \sqrt{6}\, \tilde\chi\right)
\end{eqnarray}
and for the $U(1)_Y$  gauge coupling
\begin{eqnarray} 
(1+\kappa^2) \frac{1}{a_Y}= \frac{1}{a_5}+\kappa^2 \frac{1}{a_{\chi}}
\end{eqnarray} 
or equivalently, 
\begin{eqnarray}
\frac{1}{a_Y}=\frac{1}{25}\frac{1}{a_5}+\frac{24}{25} \frac{1}{a_{\chi}} \,.
\end{eqnarray} 
For initial values $a_{\chi}=a_5$, we obtain the standard relation
of $SU(5)$.   In general, however, $a_{\chi}\ne a_5$, and there is more flexibility.

\subsection{Modular-invariant Flipped $SU(5)$}

\begin{table}[h!] 
\begin{center}
\begin{tabular}{| l |c c c|}
\hline \hline
Fields & $SU(5)\times U(1)_\chi$ & $A_4$ & $2k$\\ 
\hline \hline
$F_1 = \{Q_1, d_1^c, \nu^c_1\}$ & $(\mathbf{10}, -\frac12)$ & $\mathbf{1}$ & $+3$\\
$F_2 = \{Q_2, d_2^c, \nu^c_2\}$ & $(\mathbf{10}, -\frac12)$ & $\mathbf{1}$ & $+2$\\
$F_3 = \{Q_3, d_3^c, \nu^c_3\}$ & $(\mathbf{10}, -\frac12)$ & $\mathbf{1}$ & $+1$\\
$\bar{f} = \{u^c, L\}$ & $(\bar{\mathbf{5}}, +\frac32)$ & $\mathbf{3}$ & $-2$\\
$e^c_1 = e^c$ & $(\mathbf{1}, -\frac52)$ & $\mathbf{1}'$ & $+6$\\
$e^c_2 = \mu^c$ & $(\mathbf{1}, -\frac52)$ & $\mathbf{1}''$ & $+4$\\
$e^c_3 = \tau^c$ & $(\mathbf{1}, -\frac52)$ & $\mathbf{1}$ & $+2$\\
$\nu_S$ & $(\mathbf{1}, 0)$ & $\mathbf{3}$ & $0$\\
\hline 
$H$ & $(\mathbf{10}, -\frac12)$ & $\mathbf{1}$  & $0$\\
$\bar{H}$ & $(\overline{\mathbf{10}}, +\frac12)$ & $\mathbf{1}$  & $-1$ \\
$\Phi$ & $(\mathbf{5}, +1)$ & $\mathbf{1}$  & $0$ \\
$\tilde{\Phi}$ & $(\bar{\mathbf{5}}, -1)$ & $\mathbf{1}$ & $-1$ \\\hline 
$\xi$ & $(\mathbf{1}, 0)$ & $\mathbf{1}$ & $-1$ \\
\hline 
$Y^{2k}_{\bf r}$ & $(\mathbf{1}, 0)$ & $\mathbf{r}$ & $2k$ \\
\hline \hline
\end{tabular}
\end{center}
\caption{Transformation properties of leptons, Yukawa couplings and right-handed neutrino masses in $SU(5)\times U(1)_\chi \times A_4$, where $2k$ is the modular weight. $\bar{f}$ in the flavour space is arranged as $\bar{f} = \{ \bar{f}_1, \bar{f}_3, \bar{f}_2\}$. 
Apart from the fermions and Higgs superfields, we also include a weighton superfield $\xi$. \label{tab:particle_contents}}
\end{table}

Working in the Flipped $SU(5)$ framework, we introduce modular invariance in the flavour space with chiral matter fields arranged as multiplets of $A_4$. Assignments of matter and Higgs fields in $SU(5)\times U(1)_\chi$ and $A_4$ are shown in Table~\ref{tab:particle_contents}. In addition, a weighton $\xi$, which is a singlet scalar with non-trivial modular weight, is introduced to generate fermion mass hierarchies \cite{King:2020qaj,King:2021fhl}. 

We discuss the constraint of the modular symmetry to the flavour structure. For the up quarks, the Yukawa superpotential, as shown in Eq.~\eqref{eq:FermCoupl}, takes the form $F \bar{f} \tilde{\Phi}$. Now including the flavour indices and constraining the superpotential by the modular invariance, we obtain the most general modular-invariant superpotential terms generating quark masses
\begin{eqnarray} \label{eq:W_u_d}
{\cal W}_u &\supset& \sum_{i=1,2,3} \lambda^u_{3i} \tilde{\xi}^{3-i} Y_{\bf 3}^{(2)} F_i \bar{f} \tilde{\Phi} + \sum_{i=1,2} \lambda^u_{2i} \tilde{\xi}^{5-i} Y_{\bf 3}^{(4)} F_i \bar{f} \tilde{\Phi} + \lambda^u_{11} \tilde{\xi}^{6} Y_{{\bf 3},1}^{(6)} F_1 \bar{f} \tilde{\Phi} + \lambda^u_{12} \tilde{\xi}^{6} Y_{{\bf 3},2}^{(6)} F_1 \bar{f} \tilde{\Phi} \,, \nonumber\\
{\cal W}_d &\supset& \sum_{i,j=1,2,3} \lambda^d_{ij}  \tilde{\xi}^{8-i-j} F_i F_j \Phi  \,,
\end{eqnarray}
where $\lambda^u_{ij}$ and $\lambda^d_{ij} = \lambda^d_{ji}$ are free parameters and $\tilde{\xi} \equiv \xi /\Lambda$. {Subdominant terms such as $\tilde{\xi}^5 Y_{3}^{(6)} F_{2} \bar{f} \tilde{\Phi}$ and $\tilde{\xi}^6 Y_{3}^{(6)} F_{3} \bar{f} \tilde{\Phi}$ are possible in superpotential. However, these terms do not lead to significant deviations and thus we did not write them explicitly.}
These terms generate hierarchical Yukawa structures for quarks after the weighton $\xi$ acquires the VEV $v_\xi$. We write out $Y_d$ and $Y_u$  up to $\epsilon^6$ (with $\epsilon \equiv v_\xi/\Lambda$) as
\begin{eqnarray}\label{eq:Y_u_d}
Y_u &=& \begin{pmatrix} 
\epsilon^6 Y^{u(6)}_1+ \epsilon^4 \lambda^u_{21} Y^{(4)}_1+ \epsilon^2 \lambda^u_{31} Y_1 & \epsilon^3 \lambda^u_{22} Y^{(4)}_1+\epsilon \lambda^u_{32} Y_1 & \lambda^u_{33} Y_1 \\
\epsilon^6 Y^{u(6)}_2+ \epsilon^4 \lambda^u_{21} Y^{(4)}_2+ \epsilon^2 \lambda^u_{31} Y_2 & \epsilon^3 \lambda^u_{22} Y^{(4)}_2+\epsilon \lambda^u_{32} Y_2 &  \lambda^u_{33} Y_2 \\
\epsilon^6 Y^{u(6)}_3+ \epsilon^4  \lambda^u_{21}Y^{(4)}_3+ \epsilon^2 \lambda^u_{31} Y_3 & \epsilon^3  \lambda^u_{22} Y^{(4)}_3+\epsilon \lambda^u_{32} Y_3 & \lambda^u_{33} Y_3 \\ 
  \end{pmatrix}^\dag, \nonumber\\
Y_d &=& \begin{pmatrix} 
\lambda^d_{11} \epsilon^6 & \lambda^d_{12} \epsilon^5 & \lambda^d_{13} \epsilon^4 \\ 
\lambda^d_{12} \epsilon^5 & \lambda^d_{22} \epsilon^4 & \lambda^d_{23} \epsilon^3 \\
\lambda^d_{13} \epsilon^4 & \lambda^d_{23} \epsilon^3 & \lambda^d_{33} \epsilon^2 \end{pmatrix}^* \,,
\end{eqnarray}
where $Y_i^{(4)}$  for $i=1,2,3$ represent the three components of modular form $Y_{\bf 3}^{(4)}$ of weight $4$, and $Y_i^{u(6)}$ represent three components of the linear combination of modular forms $\lambda^u_{11} Y_{{\bf 3}_1}^{(6)} + \lambda^u_{12} Y_{{\bf 3}_2}^{(6)}$ of weight $6$. Here, we have written the Yukawa matrices in the left-right notation, where ``$^*$'' and ``$^\dag$'' represent the complex and Hermitian conjugations, respectively. $Y_u$ can be diagonalised via $Y_u = V_u\,  {\rm diag} \{ \tilde{y}_u, \tilde{y}_c, \tilde{y}_t \} \, V_u^{\prime \dag}$, where both $V_u$ and $V_u^{\prime}$ are unitary matrices. $Y_d$, as a complex and symmetric matrix, can be diagonalised via $Y_d = V_d \, {\rm diag} \{ \tilde{y}_d, \tilde{y}_s, \tilde{y}_b \} \, V_d^T$.  The quark mixing matrix is given  by $V_{\rm CKM} = V_u^\dag V_d$. 
The complexity of $Y_u$ can be further addressed in the case of small $\epsilon$. Then, $Y_u$ is approximatively written to be 
\begin{eqnarray}\label{eq:Y_u2}
Y_u &\approx& \left[
\begin{pmatrix} 
1 & \epsilon \frac{\lambda^u_{11}}{\lambda^u_{22}} & \epsilon^2 \frac{\lambda^u_{31}}{\lambda^u_{33}} \\ 
-\epsilon \frac{\lambda^u_{11}}{\lambda^u_{22}} & 1 & \epsilon \frac{\lambda^u_{32}}{\lambda^u_{33}} \\ 
-\epsilon^2 \frac{\lambda^u_{31}}{\lambda^u_{33}} & -\epsilon \frac{\lambda^u_{32}}{\lambda^u_{33}} & 1 \end{pmatrix}
\begin{pmatrix} 
\epsilon^6 Y^{(6)}_1 & \epsilon^6 Y^{(6)}_2 & \epsilon^6 Y^{(6)}_3 \\
\epsilon^3 \lambda^u_{22} Y^{(4)}_1 & \epsilon^3 \lambda^u_{22} Y^{(4)}_2 & \epsilon^3 \lambda^u_{22} Y^{(4)}_3 \\
\lambda^u_{33} Y_1 &  \lambda^u_{33} Y_2 &  \lambda^u_{33} Y_3 \end{pmatrix} \right]^* \,.
\end{eqnarray}
Eigenvalues of $Y_u$, i.e., Yukawa couplings of $u$, $c$ and $t$, can be analytically derived accordingly,
\begin{eqnarray}
\tilde{y}_u &\approx& \epsilon^6 \left[
Y^{u(6)}\cdot Y^{(6)} - \frac{|Y^{u(6)} \cdot Y^{(2)}|^2}{Y^{(2)}\cdot Y^{(2)}} \right. \nonumber\\
&& \left.
- \frac{(Y^{(2)}\cdot Y^{(2)}) (Y^{u(6)} \cdot Y^{(4)}) - (Y^{u(6)} \cdot Y^{(2)}) (Y^{(2)} \cdot Y^{(4)})}{(Y^{(2)}\cdot Y^{(2)})(Y^{(4)}\cdot Y^{(4)}) - |Y^{(4)} \cdot Y^{(2)}|^2}
\right]^{1/2} \,, \nonumber\\
\tilde{y}_c &\approx& \epsilon^3 |\lambda^u_{22}| \left[
Y^{(4)}\cdot Y^{(4)} - 
\frac{|Y^{(4)} \cdot Y^{(2)}|^2}{Y^{(2)}\cdot Y^{(2)}}
\right]^{1/2} \,, \nonumber\\
\tilde{y}_t &\approx& |\lambda^u_{33}| \sqrt{Y^{(2)}\cdot Y^{(2)}} \,,
\end{eqnarray}
where $Y^{(2)} = (Y_1, Y_2, Y_3)^T$, and the dot between two vectors $a$ and $b$ denotes the product $\sum_i a_i b_i^*$. This expression shows that Yukawa couplings of the first, second and third generation up quarks are determined by modular forms of weights $2k=6,4,2$, respectively. 

In the charged lepton sector, the superpotential terms to generate charged lepton masses are given by
\begin{eqnarray}
{\cal W}_l = \lambda_e \tilde{\xi}^6 Y_{\bf 3}^{(2)} \bar{f} e^c \Phi + \lambda_\mu \tilde{\xi}^4 Y_{\bf 3}^{(2)} \bar{f} \mu^c \Phi + \lambda_\tau \tilde{\xi}^2 Y_{\bf 3}^{(2)} \bar{f} \tau^c \Phi 
\,.
\end{eqnarray}
Here, $\lambda_e$, $\lambda_\mu$ and $\lambda_\tau$ are dimensionless coefficients which can always be kept real by rotating phases of $e^c$, $\mu^c$ and $\tau^c$. 
The Yukawa matrix is given by
\begin{eqnarray}
Y_e = 
\begin{pmatrix} 
\epsilon^6 \lambda_e Y_3 & \epsilon^4 \lambda_\mu Y_2 & \epsilon^2 \lambda_\tau Y_1 \\
\epsilon^6 \lambda_e Y_1 & \epsilon^4 \lambda_\mu Y_3 & \epsilon^2 \lambda_\tau Y_2 \\
\epsilon^6 \lambda_e Y_2 & \epsilon^4 \lambda_\mu Y_1 & \epsilon^2 \lambda_\tau Y_3 \end{pmatrix}^* \,.
\end{eqnarray}
Approximatively, the three eigenvalues are given by
\begin{eqnarray}
&&\tilde{y}_e \approx \epsilon^6 \lambda_e \left[
\frac{|Y_1^3+Y_2^3+Y_3^3 - 3 Y_1 Y_2 Y_3|^2}{|Y^{(2)}\cdot Y^{(2)}|^2 - |Y^{(2)\prime} \cdot Y^{(2)}|^2}
\right]^{1/2} \,, \nonumber\\
&&\tilde{y}_\mu \approx \epsilon^4 \lambda_\mu \left[
Y^{(2)}\cdot Y^{(2)} - 
\frac{|Y^{(2)\prime} \cdot Y^{(2)}|^2}{Y^{(2)}\cdot Y^{(2)}}
\right]^{1/2} \,, \nonumber\\
&&
\tilde{y}_\tau \approx \epsilon^2 \lambda_\tau \sqrt{Y^{(2)}\cdot Y^{(2)}} \,,
\end{eqnarray}
where $Y^{(2)\prime} = (Y_2, Y_3, Y_1)^T$. Different from the up quarks, Yukawa couplings of charged leptons are determined by only the modular forms of weight $2k=2$. 

In the neutrino sector, the Yukawa matrix $Y_D$ between $\nu$ and $\nu^c$ is obtained from the terms ${\cal W}_u$, which lead to the Yukawa matrix relation $Y_D = Y_u^T$. 
Majorana masses for $\nu^c$ are generated via 
\begin{eqnarray}
{\cal W}_{\nu^c} &=& \sum_{i,j=1,2,3} \frac{\lambda_{ij}^c}{\Lambda_c} \tilde{\xi}^{6-i-j} F_i F_j \bar{H} \bar{H} + \cdots \,,
\end{eqnarray}
where the dots represent negligible terms such as $\tilde{\xi}^{10-i-j} F_i F_j \bar{H} \bar{H}$, which are also allowed by modular invariance but further suppressed by at least $\epsilon^4$. 
${\cal W}_{\nu^c}$ leads to the Majorana mass matrix for $\nu^c$
\begin{eqnarray}
M_R &=&\frac{\langle \bar{\tilde{\nu}}^c_H \rangle^2}{\Lambda_c}
\begin{pmatrix} \lambda_{11}^c \epsilon^4 & \lambda_{12}^c \epsilon^3 & \lambda_{13}^c \epsilon^2 \\ 
\lambda_{12}^c \epsilon^3 & \lambda_{22}^c \epsilon^2 & \lambda_{23}^c \epsilon \\ \lambda_{13}^c \epsilon^2 & \lambda_{23}^c \epsilon & \lambda_{33}^c \end{pmatrix}^* \,. 
\end{eqnarray}
The light neutrino masses, after integrating out $\nu^c$, are generated via the type-I seesaw formula, i.e.,
\begin{eqnarray}
M_\nu = - M_D M_R^{-1} M_D^T\,.
\end{eqnarray}

Note that non-renormalisable superpotential terms as
\begin{eqnarray}
\sum_{i=0}^{\infty}\tilde{\xi}^{3+2i} Y_{\bf 1}^{(4+2i)} \Lambda  \Phi \tilde{\Phi}
\end{eqnarray}
cannot be forbidden by the modular symmetry. These terms, they lead to Higgs mass $\mu \sim \tilde{\xi}^3 \Lambda$ if $Y_{\bf 1}^{(4)} \neq 0$ and at some stabilisers \cite{Gui-JunDing:2019wap,deMedeirosVarzielas:2020kji}, apply, $Y_{\bf 1}^{(4)} = 0$ and may be satisfied, $\mu \sim \tilde{\xi}^5 \Lambda$. These suppressions are not enough to restrict $\mu\sim$ TeV scale.
{It is possible to introduce an additional $Z_2$ symmetry (with $\tilde{\Phi}$, $\bar{f}$ and $e^c_i$ be $Z_2$-odd and the other particles $Z_2$-even) beyond the modular symmetry which can forbid these terms and does not lead to additional corrections to the flavour structure. }

\subsection{A mechanism for incomplete representations}

As already noted,  spontaneous symmetry breaking of 
the plain field theory Flipped $SU(5)$ model  down to the SM symmetry, entails 
certain fermion mass  relations  emanating from their common origin in the original $SU(5)\times U(1)$-invariant Yukawa lagrangian. Indeed, recall that in the present  model,  the 
up-quark and Dirac neutrinos satisfy the relation $m_D^T=m_u$ at the GUT scale.
When seeking an ultra-violet completion of the model, however, this is not always true.  In string theory constructions, 
such as the heterotic and F-theory models, quite often the various GUT representations
accommodating the MSSM fields  are truncated  by stringy projection mechanisms,  magnetic 
fluxes etc~\footnote{See for example~\cite{IbanezUranga}.} , and as a result, such strict relations among the mass matrices are no longer true. In the case of F-theory constructions, in particular, this observation can be illustrated with the following  example. Within a  generic context of F-theory 
constructions, matter representations are trapped on the various intersections  of  the GUT `surface'  wrapped by the appropriate number of 7-branes, with other 7-branes
perpendicular to the GUT divisor.   As a consequence, an equal number of two-dimensional Riemann surfaces usually called `matter curves' is formed where the GUT symmetry is further enhanced.  In the simplest scenario, each one of the matter curves and
the GUT representation residing on them, are characterised by  distinct $U(1)$ `charges'   associated with the Cartan algebra of some covering group. 
For the sake of the argument, therefore,
let us assume now  that  there are $M_{10}$ copies of  chiral ten-plets on an appropriate  matter curve, i.e., $\#({\bf 10}_{-\frac 12}-\overline{\bf 10}_{\frac 12})=M_{10}$ and analogously $M_1, M_2$ copies of five-plets on two other intersections, 
$\#(\bar {\bf 5}^{(1)}_{\frac 32}-{\bf 5}^{(1)}_{-\frac 32})$, $\#(\bar {\bf 5}^{(2)}_{-\frac 32}-{\bf 5}^{(2)}_{\frac 32})$ with all of them  accommodating  fermion generations. Turing on a hypercharge flux of $N,N_1,N_2$ units respectively, we obtain:
\begin{equation}
	\mathbf{10}_i=\begin{cases}(\mathbf{3},2)_{i},\; M_{10}\\
		(\bar{\mathbf{3}},1)_{i},\;M_{10}+N\\
		(\mathbf{1},1)_{i},\;M_{10}-N
	\end{cases}\!\!\!,\;\;\;
\bar{\mathbf{5}}_1=\begin{cases}(\bar{\mathbf{3}},1)_1,\;M_1\\
		(\mathbf{1},2)_{1},\;M_1-N_1
	\end{cases}\!\!\!,\;\;\;
\bar{\mathbf{5}}_2=\begin{cases}(\bar{\mathbf{3}},1)_2,\;M_2\\
	(\mathbf{1},2)_{2},\;M_2-N_2
\end{cases}\!\!\!.
\end{equation}
%
 As an  example we take $M_{10}=3, N=0, M_1=1, M_2=0,  N_1=-N_2=1$, thence 
$\bar{\mathbf{5}}_1 \to (u^c, 0)$, $\bar{\mathbf{5}}_2 \to  (0, L)$, 
and 
\begin{eqnarray} 
\mathbf{10}_i \, \bar{\mathbf{5}}_1 \, \bar{\mathbf{5}}_h \to Q_i \, u_1^c \, h_u, \;  
\mathbf{10}_i \, \bar{\mathbf{5}}_2 \, \bar{\mathbf{5}}_h \to \nu^c_i \, L_2 \, h_u\,.
\end{eqnarray}
Thus, this way the Dirac neutrino Yukawa coupling is splitted from the up quark Yukawa coupling. 

Furthermore, it is worth to note that including multiplicities of matter field representations $F$, $\bar f$ and $e^c$ can be embedded into different $\bf 16$'s of $SO(10)$, and thus the modular-invariant Flipped $SU(5)$ model can be embedded into a modular-invariant $SO(10)$. 

In any case, we can write out the superpotential terms ${\cal W}_D$ with coefficients independent of from those in ${\cal W}_u$, 
\begin{multline}
{\cal W}_D = \sum_{i=1,2,3} \lambda^D_{3i} \tilde{\xi}^{3-i} Y_{\bf 3}^{(2)} F_i \bar{f} \tilde{\Phi} + \sum_{i=1,2} \lambda^D_{2i} \tilde{\xi}^{5-i} Y_{\bf 3}^{(4)} F_i \bar{f} \tilde{\Phi} + \lambda^D_{11} \tilde{\xi}^{6} Y_{{\bf 3},1}^{(6)} F_1 \bar{f} \tilde{\Phi}
+ \lambda^D_{12} \tilde{\xi}^{6} Y_{{\bf 3},2}^{(6)} F_1 \bar{f} \tilde{\Phi}\,.
\end{multline}
Then, we arrive at a more generalised matrix where the dimensionless coefficients could be different from those in $Y_u$, i.e., 
\begin{eqnarray}
Y_D = 
\begin{pmatrix} 
\epsilon^6 Y^{D(6)}_1+ \epsilon^4 \lambda^D_{21} Y^{(4)}_1+ \epsilon^2 \lambda^D_{31} Y_1 & \epsilon^3 \lambda^D_{22} Y^{(4)}_1+\epsilon \lambda^D_{32} Y_1 & \lambda^D_{33} Y_1 \\ 
\epsilon^6 Y^{D(6)}_2+ \epsilon^4 \lambda^D_{21} Y^{(4)}_2+ \epsilon^2 \lambda^D_{31} Y_2 & \epsilon^3 \lambda^D_{22} Y^{(4)}_2+\epsilon \lambda^D_{32} Y_2 & \lambda^D_{33} Y_2 \\
\epsilon^6 Y^{D(6)}_3+ \epsilon^4  \lambda^D_{21}Y^{(4)}_3+ \epsilon^2 \lambda^D_{31} Y_3 & \epsilon^3  \lambda^D_{22} Y^{(4)}_3+\epsilon \lambda^D_{32} Y_3 & \lambda^D_{33} Y_3  \end{pmatrix}^* \,,
\end{eqnarray} 
where $Y_i^{D(6)}$ represent the linear combination of modular forms $\lambda^D_{11} Y_{{\bf 3}_1}^{(6)} + \lambda^D_{12} Y_{{\bf 3}_2}^{(6)}$ of weight $6$. We consider an ecomonical case that the first two generations of $u^c$ and $L$ are splitted but the third generation does not. Thus, $\lambda^D_{ij}$ are independent of $\lambda^u_{ij}$ except $\lambda^D_{33}$. Although $Y_D$ is still hierarchical, the relaxing of the coefficients makes the model easier to fit the numerical data, as will be discussed in the next subsection. 
$M_\nu$ in this case is estimated to be 
\begin{eqnarray}
M_\nu \sim \begin{pmatrix} 
{\cal O}(\epsilon^4) & {\cal O}(\epsilon^3) & {\cal O}(\epsilon^2) \\
{\cal O}(\epsilon^3) & {\cal O}(\epsilon^2) & {\cal O}(\epsilon^1) \\
{\cal O}(\epsilon^2) & {\cal O}(\epsilon^1) & {\cal O}(\epsilon^0) 
\end{pmatrix}\,.
\end{eqnarray}
This kind of mass structure predicts masses $m_1: m_2: m_3 \sim \epsilon^4: \epsilon^2: 1$ and therefore, normal ordering ($m_1 < m_2 < m_3$) for neutrino masses. However, in order to generate the correct mass difference ratio $\Delta m^2_{21}/|\Delta m^2_{31}| \sim 0.03$, a fine-tuning of order $10^{-2}$ is required. 

\subsection{Numerical analysis}

We perform a $\chi^2$ analysis at the GUT scale to show how well the model fits the data.

In quark and charged lepton sectors, we apply the following  best-fit and $\pm 1 \sigma$ values as inputs of Yukawa couplings at the GUT scale
\begin{eqnarray}
&&\hspace{-4mm} \tilde{y}_u = (2.92 \pm 1.81) \times 10^{-6}, \quad\;
\tilde{y}_c = (1.43\pm0.100) \times 10^{-3}, \quad
\tilde{y}_t =  0.534 \pm 0.0341, \\
&&\hspace{-4mm} \tilde{y}_d = (4.81\pm1.06)\times10^{-6}, \quad\;
\tilde{y}_s = (9.52\pm1.03)\times10^{-5}, \quad\;\,
\tilde{y}_b =(6.95 \pm 0.175) \times 10^{-3}, \nonumber\\
&&\hspace{-4mm} \tilde{y}_e = (1.97 \pm 0.0236) \times 10^{-6}, \;
\tilde{y}_\mu = (4.16 \pm 0.0497) \times 10^{-4}, \;
\tilde{y}_\tau = (7.07 \pm 0.0727) \times 10^{-3}. \nonumber
\end{eqnarray}
These data were derived from a minimal SUSY breaking
scenario, with $\tan \beta = 5$ \cite{Antusch:2013jca,Bjorkeroth:2015ora,Okada:2019uoy,King:2020qaj}. 
They are insensitive to the exact values of $\tan \beta$ unless a very large $\tan \beta$ is taken. Three mixing angles and one CP-violating phase in the CKM mixing matrix are applied from the same literature,
\begin{eqnarray}
&\theta^q_{12} = 13.027^\circ \pm 0.0814^\circ, \;
\theta^q_{23} = 2.054^\circ \pm 0.384^\circ, \;
\theta^q_{13} = 0.1802^\circ \pm 0.0281^\circ, \; \nonumber\\
&\delta^q =69.21^\circ \pm 6.19^\circ.
\end{eqnarray}
For neutrino masses and lepton mixing, we take global best-fit values (without including SK atmospheric data) from NuFIT 5.0 \cite{Esteban:2020cvm, nufit5} and average the positive and negative $1\sigma$ errors. 
\begin{eqnarray}
&\Delta m^2_{21} = (7.42 \pm 0.21) \times 10^{-5} {\rm eV}^2 \,, \quad
\Delta m^2_{31} = (2.514 \pm 0.028) \times 10^{-3} {\rm eV}^2 \nonumber\\
&\theta_{12} = 33.44^\circ \pm 0.77^\circ\,, \quad
\theta_{23} = 49.0^\circ \pm 1.3^\circ\,, \quad
\theta_{13} = 8.57^\circ \pm 0.13^\circ\,, 
\end{eqnarray}
for the normal ordering (NO, i.e., $m_1 < m_2 < m_3$) of neutrino masses. For $\tan \beta \lesssim 10$, the RG running effect  mainly leads to an small overall enhancement to the neutrino mass scale but has negligible correction to the flavour structure due to the suppression of the charged lepton Yukawa couplings \cite{Zhou:2014sya}. Thus, in this paper, we will directly use the measured values of lepton mixing angles and mass squared differences in the numerical fit. 

\begin{table}[h!]
\begin{center}
\begin{tabular}{|c|c||c|c|}
\hline
 $\chi _q^2$ & $3.18401$ & $\chi _l^2$ & $5.06176$ \\\hline\hline
 $\lambda _{11}^u$ & $-0.767327+0.349061 i$ & $\lambda _{11}^D$ & $0.339745\, +0.27744 i$ \\\hline
 $\lambda _{12}^u$ & $0.263207\, +0.950138 i$ & $\lambda _{12}^D$ & $0.399021\, -0.508852 i$ \\\hline
 $\lambda _{21}^u$ & $-0.00316804+0.0173877 i$ & $\lambda _{21}^D$ & $0.00175927\, -0.00351132 i$ \\\hline
 $\lambda _{22}^u$ & $0.0567532\, -0.0424349 i$ & $\lambda _{22}^D$ & $0.0388083\, +0.968816 i$ \\\hline
 $\lambda _{31}^u$ & $0.0104356\, -0.0229778 i$ & $\lambda _{31}^D$ & $0.239276\, -0.906576 i$ \\\hline
 $\lambda _{32}^u$ & $-0.116958-0.00487694 i$ & $\lambda _{32}^D$ & $0.0908708\, +0.0193008 i$ \\\hline
 $\lambda _{33}^u$ & $-0.0061682-0.123591 i$ & $\lambda _{33}^D$ & $\equiv\lambda _{33}^u$ \\\hline
 $\lambda _{11}^d$ & $0.159578\, +0.124667 i$ & $\lambda _{11}^c$ & $0.283262\, +0.102305 i$ \\\hline
 $\lambda _{12}^d$ & $-1.0342+0.0286299 i$ & $\lambda _{12}^c$ & $0.102422\, +0.805828 i$ \\\hline
 $\lambda _{13}^d$ & $0.891457\, -0.105821 i$ & $\lambda _{13}^c$ & $0.35523\, -0.179145 i$ \\\hline
 $\lambda _{22}^d$ & $-0.314388-0.432018 i$ & $\lambda _{22}^c$ & $0.0475048\, +0.162506 i$ \\\hline
 $\lambda _{23}^d$ & $0.0512689\, +0.365322 i$ & $\lambda _{23}^c$ & $-0.791368-0.16351 i$ \\\hline
 $\lambda _{33}^d$ & $0.0823574\, +0.614062 i$ & $\lambda _{33}^c$ & $0.537351\, +0.829515 i$ \\\hline
 $\tau$ & $1.48709 + 0.310071 i$ & $\lambda _e$ & $0.343493$ \\\hline
 $\epsilon$ & $0.10566$ & $\lambda _{\mu }$ & $0.805984$ \\\hline
 \multicolumn{2}{|c||}{}  & $\lambda _{\tau }$ & $0.15144$ \\\hline\hline
 $\tilde{y}_d$ & $2.9996 \times 10^{-6}$ & $\tilde{y}_e$ & $1.974 \times 10^{-6}$ \\\hline
 $\tilde{y}_s$ & $0.0000958$ & $\tilde{y}_{\mu }$ & $0.0004173$ \\\hline
 $\tilde{y}_b$ & $0.0069456$ & $\tilde{y}_{\tau }$ & $0.0070656$ \\\hline
 $\tilde{y}_u$ & $2.7243 \times 10^{-6}$ & $m_1$ & $0.00086566$~eV \\\hline
 $\tilde{y}_c$ & $0.00143656$ & $m_2$ & $0.00866672$~eV \\\hline
 $\tilde{y}_t$ & $0.519732$ & $m_3$ & $0.0501839$~eV \\\hline
 $\theta _{12}^q$ & $13.0452^\circ$ & $\theta _{12}$ & $33.272^\circ$ \\\hline
 $\theta _{23}^q$ & $2.06442^\circ$ & $\theta _{23}$ & $46.1526^\circ$ \\\hline
 $\theta _{13}^q$ & $0.17796^\circ$ & $\theta _{13}$ & $8.60762^\circ$ \\\hline
 $\delta ^q$ & $68.4553^\circ$ & $\delta $ & $-7.97559^\circ$ \\\hline
 \multicolumn{2}{|c||}{} & $M_1$ & $4.66397\times 10^8$~GeV \\\cline{3-4}
 \multicolumn{2}{|c||}{} & $M_2$ & $1.25924\times 10^{11}$~GeV \\\cline{3-4}
 \multicolumn{2}{|c||}{} & $M_3$ & $1.5415\times 10^{13}$~GeV \\\hline
\end{tabular}
\caption{\label{tab:bm1} Inputs and predictions in Benchmark 1.}
\end{center}
\end{table}

\begin{table}[h!]
\begin{center}
\begin{tabular}{|c|c||c|c|}
\hline
 $\chi _q^2$ & $8.55801$ & $\chi _l^2$ & $1.24054$ \\\hline\hline
 $\lambda _{11}^u$ & $-0.23631+0.0206202 i$ & $\lambda _{11}^D$ & $-0.0370347-0.133943 i$ \\\hline
 $\lambda _{12}^u$ & $-0.457106-0.212065 i$ & $\lambda _{12}^D$ & $0.00890465\, -0.0645904 i$ \\\hline
 $\lambda _{21}^u$ & $0.0547413\, -0.00985699 i$ & $\lambda _{21}^D$ & $-0.587964-0.698841 i$ \\\hline
 $\lambda _{22}^u$ & $0.20306\, -0.213503 i$ & $\lambda _{22}^D$ & $-1.5934+3.98198 i$ \\\hline
 $\lambda _{31}^u$ & $0.0535951\, -0.00835941 i$ & $\lambda _{31}^D$ & $0.032806\, +0.101563 i$ \\\hline
 $\lambda _{32}^u$ & $0.143194\, -0.0289485 i$ & $\lambda _{32}^D$ & $-0.137639+0.0403966 i$ \\\hline
 $\lambda _{33}^u$ & $0.147185\, -0.213406 i$ & $\lambda _{33}^D$ & $\equiv\lambda _{33}^u$ \\\hline\hline
 $\lambda _{11}^d$ & $0.872726\, -0.199401 i$ & $\lambda _{11}^c$ & $-0.258699-0.149259 i$ \\\hline
 $\lambda _{12}^d$ & $-0.022182-0.107737 i$ & $\lambda _{12}^c$ & $0.644902\, -0.835538 i$ \\\hline
 $\lambda _{13}^d$ & $0.555058\, +0.563634 i$ & $\lambda _{13}^c$ & $0.786701\, +0.656483 i$ \\\hline
 $\lambda _{22}^d$ & $0.165182\, -0.00517376 i$ & $\lambda _{22}^c$ & $0.00748405\, -0.161169 i$ \\\hline
 $\lambda _{23}^d$ & $0.0328347\, +0.434846 i$ & $\lambda _{23}^c$ & $0.0404458\, +0.229882 i$ \\\hline
 $\lambda _{33}^d$ & $0.224434\, +0.415195 i$ & $\lambda _{33}^c$ & $-0.0466412+0.0379145 i$ \\\hline
 $\tau$ & $0.653628\, +1.25817 i$ & $\lambda _e$ & $0.340752$ \\\hline
 $\epsilon$ & $0.121925$ & $\lambda _{\mu }$ & $0.971434$ \\\hline
  \multicolumn{2}{|c||}{} & $\lambda _{\tau }$ & $0.225573$ \\\hline\hline
 $\tilde{y}_d$ & $3.40662 \times 10^{-6}$ & $\tilde{y}_e$ & $1.96672 \times 10^{-6}$ \\\hline
 $\tilde{y}_s$ & $0.000101448$ & $\tilde{y}_{\mu }$ & $0.000413715$ \\\hline
 $\tilde{y}_b$ & $0.00679233$ & $\tilde{y}_{\tau }$ & $0.00705526$ \\\hline
 $\tilde{y}_u$ & $3.67168 \times 10^{-6}$ & $m_1$ & $0.00086566$~eV \\\hline
 $\tilde{y}_c$ & $0.00140942$ & $m_2$ & $0.00866672$~eV \\\hline
 $\tilde{y}_t$ & $0.571741$ & $m_3$ & $0.0501839$~eV \\\hline
 $\theta _{12}^q$ & $13.0549^\circ$ & $\theta _{12}$ & $33.8375^\circ$ \\\hline
 $\theta _{23}^q$ & $2.37476^\circ$ & $\theta _{23}$ & $49.9436^\circ$ \\\hline
 $\theta _{13}^q$ & $0.204838^\circ$ & $\theta _{13}$ & $8.5851^\circ$ \\\hline
 $\delta ^q$ & $79.1935^\circ$ & $\delta $ & $-7.97559^\circ$ \\\hline
 \multicolumn{2}{|c||}{} & $M_1$ & $2.26121\times 10^{10}$~GeV \\\cline{3-4}
 \multicolumn{2}{|c||}{} & $M_2$ & $2.21381\times 10^{11}$~GeV \\\cline{3-4}
 \multicolumn{2}{|c||}{} & $M_3$ & $1.24968\times 10^{12}$~GeV \\\hline
\end{tabular}
\caption{\label{tab:bm2} Inputs and predictions in Benchmark 2.}
\end{center}
\end{table}

\begin{figure}[h!]
\centering
\includegraphics[width=1\textwidth]{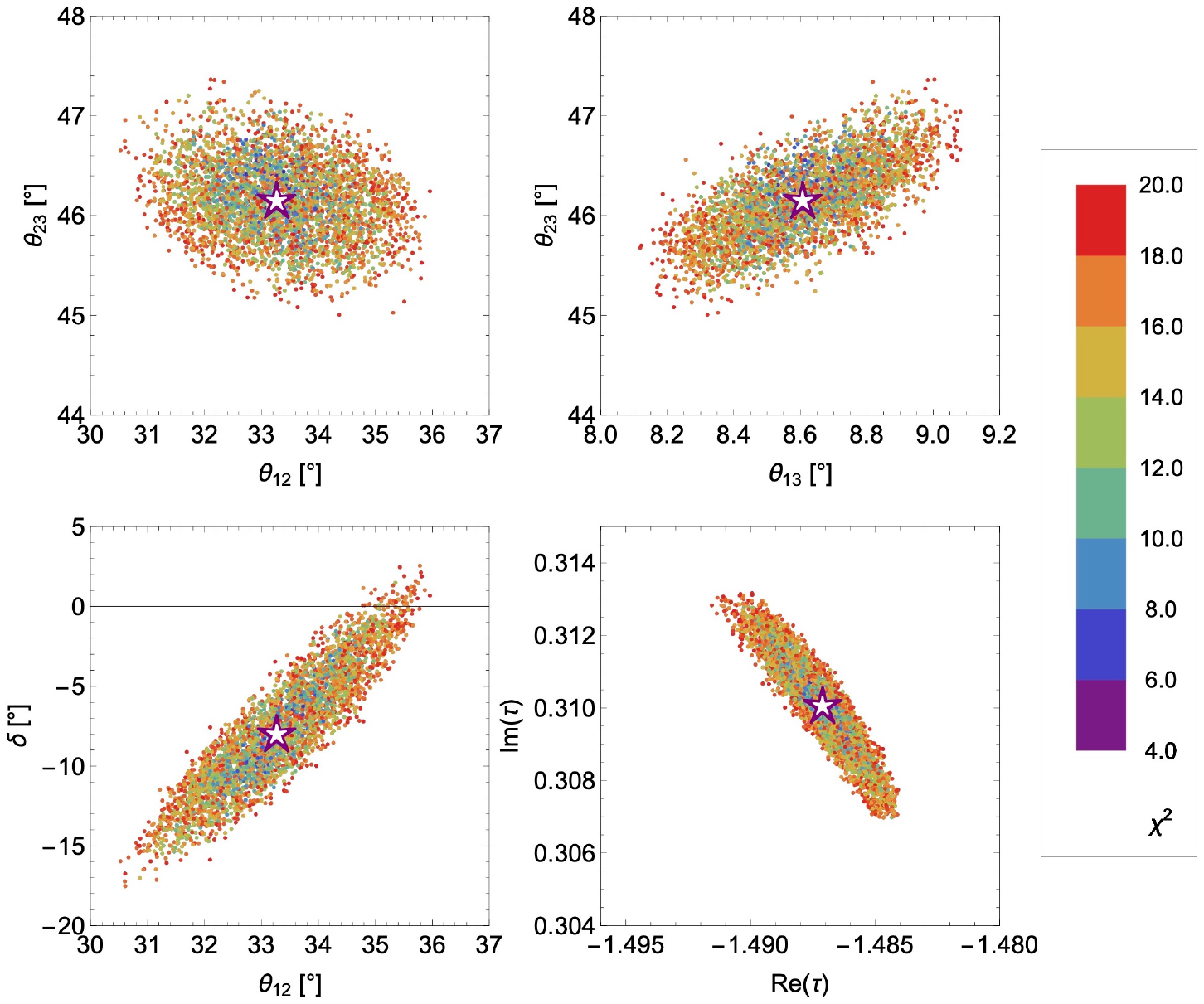}
\caption{\label{fig:predictions} Predictions of lepton mixing angles and the Dirac CP-violating oscillation phase from the scan around the first benchmark, where stars refer to predictions in the first benchmark. }
\end{figure}

We define the following two $\chi^2$ functions  in the quark sector and lepton sector, respectively,
\begin{eqnarray} \label{eq:chi}
\chi^2_q &=& \sum_{i \in O_q} \Big(\frac{p_i(P_q) - b_i}{\sigma_i}\Big)^2 \,, \nonumber\\
\chi^2_l &=& \sum_{i \in O_l} \Big(\frac{p_i(P_l) - b_i}{\sigma_i}\Big)^2 \,, 
\end{eqnarray}

{where $p_{i}$ are the model predictions, $b_{i}$ the current best-fit values and the errors $\sigma_{i}$ correspond here to the average of the $1 \sigma$ ranges for each observable.}

The relevant free parameters and observables are listed in sets
\begin{eqnarray} \label{eq:quark_set}
P_q &=& \{ \lambda^u_{11}, \lambda^u_{12}, \lambda^u_{21}, \lambda^u_{22}, \lambda^u_{31}, \lambda^u_{32}, \lambda^u_{33}, \lambda^d_{11}, \lambda^d_{12}, \lambda^d_{13}, \lambda^d_{22}, \lambda^d_{23}, \lambda^d_{33}, 
\epsilon, \tau \} \,,\nonumber\\
O_q &=& \{ \tilde{y}_u, \tilde{y}_c, \tilde{y}_t, \tilde{y}_d, \tilde{y}_s, \tilde{y}_b, \theta_{12}^q, \theta_{23}^q,  \theta_{13}^q, \delta^q\} \,,
\end{eqnarray}
and
\begin{eqnarray} \label{eq:lepton_set}
P_l &=& \{ \lambda^D_{11}, \lambda^D_{12}, \lambda^D_{21}, \lambda^D_{22}, \lambda^D_{31}, \lambda^D_{32}, \lambda^D_{33}, \lambda_e, \lambda_\mu, \lambda_\tau, \lambda^c_{11}, \lambda^c_{12}, \lambda^c_{13}, \lambda^c_{22}, \lambda^c_{23}, \lambda^c_{33}, \epsilon, \tau \} \,,\nonumber\\
O_l &=& \{ \tilde{y}_e, \tilde{y}_\mu, \tilde{y}_\tau, \Delta m^2_{21},  \Delta m^2_{31}, \theta_{12}, \theta_{23},  \theta_{13}\} \,,
\end{eqnarray}
respectively. 
Here, we ignore the contribution of $\nu_S$ to the neutrino masses. All coefficients in $P_q$ and $P_l$, i.e., $\lambda_{ij}^u$, $\lambda_{ij}^d$, $\lambda_{ij}^D$, $\lambda_{ij}^c$, and $\lambda_{e,\mu,\tau}$, are scanned with absolute values in the range $(0,1)$ with arbitrary phases in the range $(0, 2 \pi)$. In the set $P_q$ and $P_l$, same values of $\epsilon$ and $\tau$ are respectively used. $\epsilon$ is scanned in $(0, 0.2)$ and $\tau$ is  scanned in the fundamental domain of $\bar{\Gamma}(3)$. The latter is achieved by firstly scanned in the fundamental domain of $\Gamma$ and then shifted to the rest region following modular transformations of the finite modular group $\Gamma_3$, i.e., $\tau \to \gamma \tau$ for all $\gamma$ in $\Gamma_3$. Furthermore, $\lambda_{33}^D \equiv \lambda_{33}^u$ is always fixed in the scan as discussed before.

Due to the large pararemter space, a full scan in all parameter space is hard to be performed. Instead, we listed two benchmark points in Tables~\ref{tab:bm1} and \ref{tab:bm2}, respectively, as representatives. Both benchmarks fit the numerical values very well $\chi^2_q + \chi^2_l <10$. In both cases, the second octant of $\theta_{23}$, i.e., $\theta_{23}>45^\circ$ is predicted and the leptonic CP-violation is small $|\delta| < 10^\circ$. In the second benchmark, hierarchical right-handed neutrinos masses with the lightest one $M_1 \sim 2 \times 10^{10}$ GeV is predicted. It can be applied for baryogenesis via thermal leptogenesis.\footnote{If the neutrino is a Majorana fermion, the neutrino and the antineutrino are the same particle. In this case, it becomes possible to convert it
matter to antimatter and vice versa. Therefore, the existence of neutrino masses makes
possible to create an imbalance between matter and anti-matter in the early universe. A successful thermal leptogenesis requires $M_1 \gtrsim 10^9$~GeV.}
This benchmark has a relatively small coefficient $|\lambda^u_{33}| \sim 0.1$ as an input to predict the Top Yukawa coupling $\tilde{y}_t \sim 0.5$. It is achieved due to the enhancement of a large value of the modular form $Y^{(2)} = (-0.0405235-0.0260606 i,0.205766\, +0.397288 i,0.180747\, -4.15085 i)^T$ at $\tau = -1.48709 + 0.310071 i$ (remind that $\tilde{y}_t \approx |\lambda^u_{33}| \sqrt{Y^{(2)}\cdot Y^{(2)}}$). We further make a scan around the first benchmark with all free parameters deviated by less than $1\%$. Predictions of mixing angles and $\delta$ are shown in Fig.~\ref{fig:predictions}.

\section{Conclusion}

In this work, the long-standing problem of the origin of flavour mixing among the fermion families has been investigated within the framework 
of modular symmetries. Inspired by a wide class of string theory effective models endowed with modular invariance and simplicity of the Higgs sector 
to implement the symmetry breaking, we have  constructed a Flipped $SU(5)$ model with $A_4$ modular symmetry, assigning specific modular weights to  the fermion and Higgs fields. 
In this context, Yukawa couplings are modular forms, with the three families (anti-)five-plets of $SU(5)$ transforming as a triplet under the $A_4$ modular symmetry, while the fermion ten-plets transform as singlet $A_4$ representations, distinguished by different modular weights, and the charged lepton electroweak singlets transform as non-trivial one-dimensional $A_4$ representations with different modular weights. The hierarchy of charged fermion masses is then achieved via higher dimensional operators coupled to a singlet weighton field which carries unit modular weight.  

Flipped $SU(5)$ models exhibit many interesting features which make them attractive string motivated candidates as compared to standard $SU(5)$ GUTs.  Among other merits, 
only a pair of ${\bf 10}+\overline{\bf 10}$ Higgs representations suffices to break the GUT symmetry.  At the same time the down-type colour triplets of this Higgs  pair combine with those of ${\bf 5}+\bar{\bf 5}$ Higgs representations to realise  the doublet-triplet splitting  in an elegant manner. As for
the mass matrix textures, because charged right-handed lepton  fields  are $SU(5)$ singlets, the
charged lepton mass matrix is unrelated to that of the down quarks. This way the  restrictive mass constraints   of the ordinary $SU(5)$ are avoided and
modular invariance is the main symmetry left over to organise the charged fermion
mass matrices. Regarding the neutral fermion sector, a notable property, not shared by the standard $SU(5)$, is that the  right-handed neutrinos are contained in the ten-plet representation together with the quark doublets and
the down-type colour triplets. This implies the relation  $Y_D = Y_u^T$ between Dirac neutrino  and up quark Yukawa
matrices and, in the simplest version of the model, this restrictive relation makes it difficult to fit the neutrino oscillation data. 

In order to overcome this problem and avoid fine tuning issues, we appeal to a string-inspired mechanism according to which magnetic fluxes turned on along the Abelian subgroup of $SU(5)$ split the $SU(5)$-representations and disentangle the neutrino and up quark Yukawa couplings. Notwithstanding this mechanism, the Dirac neutrino mass matrix has a 
hierarchical structure, which can be partially canceled by the hierarchical Majorana mass matrix for right-handed neutrinos,
resulting in a normally ordered and hierarchical pattern of light neutrino masses, with $m_1$ an order of magnitude smaller than $m_2$. This way, we are able to fit the neutrino oscillation data, only with a slight fine tuning of parameters in the neutrino sector. 
For the considered model, we find a good fit to charged quark and lepton masses, with a relatively low value of $\chi^2$.
The leptonic CP-violating oscillation phase is predicted to be $\delta = -8^\circ\pm 8^\circ$.
A by-product of our approach is the prediction of hierarchical heavy neutrino masses. The lightest one may around $10^{10}$~GeV, which is around the correct value for standard thermal leptogenesis, which however we do not pursue further here.

\subsection*{Acknowledgements}
SFK acknowledges the STFC Consolidated Grant ST/L000296/1 and the European Union Horizon 2020 Research and Innovation programme under Marie Sklodowska-Curie grant agreement HIDDeN European ITN project (H2020-MSCA-ITN-2019//860881-HIDDeN).


\appendix

\section{Contribution of singlet neutrinos}

At this point, we will study the case where, we take into account the presence of additional singlet neutrino superfields $\nu_{S}$, as predicted string derived Flipped $SU(5)$ models. In this case, more neutrinos mass terms exist,
\begin{eqnarray}
{\cal W} &\supset& \sum_{i=1,2,3} \lambda^S_{3i} \tilde{\xi}^{3-i} Y_{\bf 3}^{(2)} F_{i} \nu_{S} \bar{H} + \sum_{i=1,2} \lambda^S_{2i} \tilde{\xi}^{5-i} Y_{\bf 3}^{(4)} F_i \nu_S \bar{H} 
\nonumber\\
&&  + \lambda_{11}^{S} \tilde{\xi}^{6} Y_{\bold{3},1}^{(6)} F_{1} \nu_{S} \bar{H} + \lambda_{12}^{S} \tilde{\xi}^{6} Y_{\bold{3},2}^{(6)} F_{1} \nu_{S} \bar{H}+ m_{S} \nu_{S} \nu_{S} \,.
\end{eqnarray}
These terms, together with those in ${\cal W}_u$ superpotential, generate a general $9\times 9$ mass matrix for neutrinos. In the basis $(\nu, \nu^c, \nu_S)$, this mass matrix is written as
\begin{eqnarray}
\mathcal{M}_\nu^{9 \times 9} = \begin{pmatrix} 
\mathbf{0} & M_D & \mathbf{0} \\
M_D^T & M_R & M_D' \\
\mathbf{0} & M_D^{\prime T} & M_S
\end{pmatrix}\,,
\end{eqnarray}
where all sub-blocks are $3\times 3$ matrices. In particular,  
\begin{eqnarray}\label{eq:M_S}
M_D' &=& \langle \bar{\tilde{\nu}}^c_H \rangle \!
\begin{pmatrix} 
\epsilon^6 Y^{S(6)}_1+ \epsilon^4 \lambda^S_{21} Y^{(4)}_1+ \epsilon^2 \lambda^S_{31} Y_1 & \epsilon^3 \lambda^S_{22} Y^{(4)}_1+\epsilon \lambda^S_{32} Y_1 & \lambda^S_{33} Y_1 \\
\epsilon^6 Y^{S(6)}_2+ \epsilon^4 \lambda^S_{21} Y^{(4)}_2+ \epsilon^2 \lambda^S_{31} Y_2 & \epsilon^3 \lambda^S_{22} Y^{(4)}_2+\epsilon \lambda^S_{32} Y_2 & \lambda^S_{33} Y_2 \\
\epsilon^6 Y^{S(6)}_3+ \epsilon^4  \lambda^S_{21}Y^{(4)}_3+ \epsilon^2 \lambda^S_{31} Y_3 & \epsilon^3  \lambda^S_{22} Y^{(4)}_3+\epsilon \lambda^S_{32} Y_3 & \lambda^S_{33} Y_3 \\ 
\end{pmatrix}^\dag , \nonumber\\
M_S &=& m_S
\begin{pmatrix} 1 & 0 & 0 \\ 0 & 0 & 1 \\ 0 & 1 & 0 \end{pmatrix} \,,
\end{eqnarray}
and $Y^{S(6)}_i$ for $i=1,2,3$ are three components of $\lambda^S_{11} Y_{{\bf 3},1}^{(6)} + \lambda^S_{12} Y_{{\bf 3},2}^{(6)}$. 
The light neutrino mass matrix in this case is modified into \cite{Zhou:2012ds}
\begin{eqnarray}
M_\nu = - M_D \left( M_R - M_D' M_S^{-1} M_D^{\prime T} \right)^{-1} M_D^T \,.
\end{eqnarray}
Without considering the flavour structure and assuming order-one coefficients, from the relation that applies to the double seesaw formula, $ - M_D \left( M_R - M_D' M_S^{-1} M_D^{\prime T} \right)^{-1} M_D^T $, the heaviest eigenvalues of $M_R$ and $M_D' M_S^{-1} M_D^{\prime T}$ are of order $\langle \bar{\tilde{\nu}}^c_H \rangle^2 /\Lambda_c$ and $\langle \bar{\tilde{\nu}}^c_H \rangle^2 /m_S$, respectively. Therefore, the light neutrino masses are a competition of the two scales $\Lambda_c$ and $m_S$. 

In our numerical analysis as discussed in section 3.4, we have focused only in the scenario $m_S \gg \Lambda_c \sim 10^{16}$~GeV. 
Below, we give a brief discussion on the opposite scenario $m_S \ll \Lambda_c$. Namely, the light neutrino masses is given by $M_\nu = M_D \left( M_D' M_S^{-1} M_D^{\prime T} \right)^{-1} M_D^T$, which is also called the double seesaw formula. 
At leading order of $\epsilon$, one can check that $M_\nu$ approximates to 
\begin{eqnarray}
M_\nu \sim \epsilon^{-4} \frac{m_S v_u^2}{\langle \bar{\tilde{\nu}}^c_H \rangle^2} 
\left(
\begin{array}{ccc}
 Y_{1}^2 & Y_{1} Y_{2} & Y_{1} Y_{3} \\
 Y_{1} Y_{2} & Y_{2}^2 & Y_{2} Y_{3} \\
 Y_{1} Y_{3} & Y_{2} Y_{3} & Y_{3}^2 \\
\end{array}
\right)^*
\end{eqnarray}
up to an overall factor. It partially determines the flavour structure, but has only one non-zero eigenstate $\sim \epsilon^{-4} \frac{m_S v_u^2}{\langle \bar{\tilde{\nu}}^c_H \rangle^2}$. Including the next-to-leading correction, one obtain the second lightest neutrino has mass $\frac{m_S v_u^2}{\langle \bar{\tilde{\nu}}^c_H \rangle^2}$. Their hierarchy is too large to explain the ratio $|r| \equiv \Delta m^2_{21}/|\Delta m^2_{31}| \sim 0.03$ unless fine tuning between coefficients are considered. 

\section{The second model}

In this appendix we present a second Flipped $SU(5)$ model which differs from the first with respect to the representations of the $A_{4}$ symmetry and the modular weights assigned to the fields.
The transformation properties of the spectrum are shown in Table \ref{tab:particle_contents2}.
\vspace{0.5cm}

\begin{table}[h!] 
\begin{center}
\begin{tabular}{| l |c c c|}
\hline \hline
Fields & $SU(5)\times U(1)_\chi$ & $A_4$ & $2k$\\ 
\hline \hline
$F_1 = \{Q_1, d_1^c, \nu^c_1\}$ & $(\mathbf{10}, -\frac12)$ & $\mathbf{1}$ & $+1$\\
$F_2 = \{Q_2, d_2^c, \nu^c_2\}$ & $(\mathbf{10}, -\frac12)$ & $\mathbf{1'}$ & $-3$\\
$F_3 = \{Q_3, d_3^c, \nu^c_3\}$ & $(\mathbf{10}, -\frac12)$ & $\mathbf{1''}$ & $-4$\\
$\bar{f} = \{u^c, L\}$ & $(\bar{\mathbf{5}}, +\frac32)$ & $\mathbf{3}$ & $-4$\\
$e^c_1 = e^c$ & $(\mathbf{1}, -\frac52)$ & $\mathbf{1}''$ & $+4$\\
$e^c_2 = \mu^c$ & $(\mathbf{1}, -\frac52)$ & $\mathbf{1}$ & $+3$\\
$e^c_3 = \tau^c$ & $(\mathbf{1}, -\frac52)$ & $\mathbf{1}'$ & $+1$\\
$\nu_S$ & $(\mathbf{1}, 0)$ & $\mathbf{3}$ & $0$\\
\hline 
$H$ & $(\mathbf{10}, -\frac12)$ & $\mathbf{1}'$  & $0$\\
$\bar{H}$ & $(\overline{\mathbf{10}}, +\frac12)$ & $\mathbf{1}$  & $-2$ \\
$\Phi$ & $(\mathbf{5}, +1)$ & $\mathbf{1}'$  & $+1$ \\
$\tilde{\Phi}$ & $(\overline{\mathbf{5}},-1)$ & $\mathbf{1}$ & $+4$ \\\hline 
$\xi$ & $(\mathbf{1}, 0)$ & $\mathbf{1}$ & $-1$ \\
\hline 
$Y^{2k}_{\bf r}$ & $(\mathbf{1}, 0)$ & $\mathbf{r}$ & $2k$ \\
\hline \hline
\end{tabular}
\end{center}
\caption{The representations of the second model and their transformation properties under $SU(5)\times U(1)_\chi \times A_4$.\label{tab:particle_contents2}}
\end{table}

The superpotential Yukawa couplings for the up and down quark are,
\begin{eqnarray}
	\label{sup2}
{\cal W}_u&=& \lambda_{1}^{u} F_{1} \tilde{\Phi} \bar{f} Y_{\bold{3}}^{(4)} \tilde{\xi}^{5} + \lambda_{2}^{u} F_{2} \tilde{\Phi} \bar{f} Y_{\bold{3}}^{(4)} \tilde{\xi} + \lambda_{3}^{u} F_{3} \tilde{\Phi} \bar{f} Y_{\bold{3}}^{(4)}~,
\nonumber\\
{\cal W}_d&=&\lambda_{12}^{d} F_{1} F_{2} \Phi Y_{\bold{1'}}^{(4)} \tilde{\xi}^{3} + \lambda_{13}^{d} F_{1} F_{3} \Phi Y_{\bold{1}}^{(4)} \tilde{\xi}^{2} \nonumber\\
&&+ \lambda_{22}^{d} F_{2} F_{2} \Phi Y_{\bold{1}}^{(8)} \tilde{\xi}^{3} + \lambda_{23}^{d} F_{2} F_{3} \Phi Y_{\bold{1''}}^{(8)} \tilde{\xi}^{2} + \lambda_{33}^{d} F_{3}F_{3} \Phi Y_{\bold{1'}}^{(8)} \tilde{\xi}~,
\end{eqnarray}
where $\lambda^{u}_{i}$ and $\lambda_{ij}^{d}$ are free parameters and $\tilde{\xi} \equiv \xi/\Lambda$, with $\Lambda$ a dimensionful cut-off flavour scale.  The corresponding Yukawa matrices $Y_{u}$ and $Y_{d}$ are
\begin{eqnarray}
Y_{u}&=&
\begin{pmatrix}
\lambda_{1}^{u} Y_{1}^{(4)} \tilde{\xi}^{5} & \lambda_{2}^{u} Y_{3}^{(4)} \tilde{\xi} & \lambda_{3}^{u} Y_{2}^{(4)} \\
\lambda_{1}^{u} Y_{3}^{(4)} \tilde{\xi}^{5} & \lambda_{2}^{u} Y_{2}^{(4)} \tilde{\xi} & \lambda_{3}^{u} Y_{1}^{(4)} \\
\lambda_{1}^{u} Y_{2}^{(4)} \tilde{\xi}^{5} & \lambda_{2}^{u} Y_{1}^{(4)} \tilde{\xi} & \lambda_{3}^{u} Y_{3}^{(4)} \\
\end{pmatrix}^{\dagger} \,,\nonumber\\
Y_{d}&=&
\begin{pmatrix}
0 & \lambda_{12}^{d} \hspace{0.08cm} Y_{\bold{1'}}^{(4)} \tilde{\xi}^{3} & \lambda_{13}^{d} \hspace{0.08cm} Y_{\bold{1}}^{(4)} \tilde{\xi}^{2} \\
\lambda_{12}^{d} \hspace{0.08cm} Y_{\bold{1'}}^{(4)} \tilde{\xi}^{3} & \lambda_{22}^{d} \hspace{0.08cm} Y_{\bold{1}}^{(8)} \tilde{\xi}^{3} & \lambda_{23}^{d} Y_{\bold{1''}}^{(8)} \tilde{\xi}^{2} \\
\lambda_{13}^{d} \hspace{0.08cm} Y_{\bold{1}}^{(4)} \tilde{\xi}^{2} & \lambda_{23}^{d} Y_{\bold{1''}}^{(8)} \tilde{\xi}^{2} & y_{33}^{d} Y_{\bold{1'}}^{(8)} \hspace{0.08cm} \tilde{\xi} \\
\end{pmatrix}^{*} \,,
\end{eqnarray}
where $Y_{i}^{(4)}$ for $i=1,2,3$ represent the three components of modular form $Y_{\bold{3}}^{(4)}$ of weight 4 and $Y_{{\bf r}}^{(2k)}$ are modular forms with weights $2k=4,6,8$ and the corresponding representations of the $A_{4}$ group, ${\bf r}={\bf 1}, {\bf 1'}, {\bf 1''}$. 

In the charged lepton sector, the superpotential terms generating the charged lepton masses,
\begin{equation}
{\cal W}_l = \lambda^{e} \bar{f} \Phi e^{c} Y_{\bold{3}}^{(4)} \tilde{\xi}^{5} + \lambda^{\mu} \bar{f} \Phi \mu^{c} Y_{\bold{3}}^{(4)} \tilde{\xi}^{3} + \lambda_{\tau} \bar{f} \Phi \tau^{c} Y_{\bold{3}}^{(4)} \tilde{\xi}^{2} \,,
\end{equation}
where, $\lambda^{e}$, $\lambda^{\mu}$, $\lambda^{\tau}$ are dimensionless  coefficients. So, we have the matrix,
\begin{equation}
Y_{l} =
\begin{pmatrix}
\lambda^{e} Y_{1}^{(4)} \tilde{\xi}^{5} & \lambda_{\mu} Y_{3}^{(4)} \tilde{\xi}^{4} & \lambda_{\tau} Y_{2}^{(4)} \tilde{\xi}^{2} \\
\lambda_{e} Y_{3}^{(4)} \tilde{\xi}^{5} & \lambda_{\mu} Y_{2}^{(4)} \tilde{\xi}^{4} & \lambda_{\tau} Y_{1}^{(4)} \tilde{\xi}^{2} \\
\lambda_{e} Y_{2}^{(4)} \tilde{\xi}^{5} & \lambda_{\mu} Y_{1}^{(4)} \tilde{\xi}^{4} & \lambda_{\tau} Y_{3}^{(4)} \tilde{\xi}^{2}
\end{pmatrix} \,.
\end{equation}
The superpotential terms in the neutrino sector are 
\begin{eqnarray}
{\cal W}_\nu &=&
\lambda_{1}^{u'} F_{1} \tilde{\Phi} \bar{f} Y_{\bold{3}}^{(4)} \tilde{\xi}^{5} + \lambda_{2}^{u'} F_{2} \tilde{\Phi} \bar{f} Y_{\bold{3}}^{(4)} \tilde{\xi} + \lambda_{3}^{u'} F_{3} \tilde{\Phi} \bar{f} Y_{\bold{3}}^{(4)} + \lambda_{1}^{H} F_{1} \bar{H} \nu_{S} Y_{\bold{3}}^{(6)} \tilde{\xi}^{5} 
\nonumber\\
&&+\lambda_{1}^{H '} F_{1} \bar{H} \nu_{S} Y_{\bold{3}}^{(4)} \tilde{\xi}^{3} + \lambda_{2}^{H} F_{2} \bar{H} \nu_{S} Y_{\bold{3}}^{(6)} \tilde{\xi} + \lambda_{3}^{H}  F_{3} \bar{H} \nu_{S} Y_{\bold{3}}^{(6)}
+ M_{S} \nu_{S} \nu_{S}  \,.
\end{eqnarray}
From this superpotential, we find the matrices,
\begin{eqnarray}
Y_{D}&=&\dfrac{y_{u} \hspace{0.04cm} \upsilon_{u}}{\sqrt{2}}
\begin{pmatrix}
\lambda_{1}^{u} Y_{1}^{(4)} \tilde{\xi}^{5} & \lambda_{2}^{u'} Y_{3}^{(4)} \tilde{\xi} & \lambda_{3}^{u} Y_{2}^{(4)} \\
\lambda_{1}^{u} Y_{3}^{(4)} \tilde{\xi}^{5} & \lambda_{2}^{u'} Y_{2}^{(4)} \tilde{\xi} & \lambda_{3}^{u} Y_{1}^{(4)} \\
\lambda_{1}^{u} Y_{2}^{(4)} \tilde{\xi}^{5} & \lambda_{2}^{u'} Y_{1}^{(4)} \tilde{\xi} & \lambda_{3}^{u} Y_{3}^{(4)} \\
\end{pmatrix}^{*} \,,\nonumber\\
M'_{D}&=&\langle \bar{\tilde{\nu}}^c_H \rangle
\begin{pmatrix}
\lambda_{1} Y_{1}^{(6)H} \tilde{\xi}^{5}+ \lambda_{1}^{'} {Y_{1}^{(4)H}} \tilde{\xi}^{3} & \lambda_{2} Y_{3}^{(6)H} \tilde{\xi} & \lambda_{3} {Y_{2}^{(6)H}} \\
\lambda_{1} {Y_{3}^{(6)H}} \tilde{\xi}^{5} + \lambda_{1}^{'} {Y_{3}^{(4)H}} \tilde{\xi}^{3} & \lambda_{2} {Y_{2}^{(6)H}} \tilde{\xi} & \lambda_{3} {Y_{1}^{(6)H}} \\
\lambda_{1} {Y_{2}^{(6)H}} \tilde{\xi}^{5} + \lambda_{1}^{'} Y_{2}^{(4)H} \tilde{\xi}^{3} & \lambda_{2} {Y_{1}^{(6)H}} \tilde{\xi}  & \lambda_{3} {Y_{3}^{(6)H}} \\
\end{pmatrix}^{\dag} \,,\nonumber\\
M_{S}&=&m_{S}
\begin{pmatrix}
1 & 0 & 0 \\
0 & 0 & 1 \\
0 & 1 & 0 \\
\end{pmatrix}\,, 
\label{eq:neutrino_model_II}
\end{eqnarray}
where $Y_{r}^{(2k)H}$ with $r=1,2,3$ and $2k=4,6$ represent three components of the linear combination of modular forms $Y_{{\bf 3}_{1}}^{(2k)}+Y_{{\bf 3}_{2}}^{(2k)}$.

In this model, we consider case that, only the second generation of $u^{c}$ and $L$ is splitted but the first and third generations do not. So, as we see in Eq.~\eqref{eq:neutrino_model_II} the free parameters $\lambda_{1}^{u}$ and $\lambda_{3}^{u}$ are the same as those for the domain of up quarks, while only the second parameter is different.

Also, Majorana masses for $\nu^{c}$ are generated via
\begin{equation}
\lambda_{ij}^{\nu} F_{i} F_{j} \bar{H} \bar{H} Y_{\bold{r}}^{(2k)} \tilde{\xi}^{n} \,,
\end{equation}
where $\lambda_{ij}^{\nu}$ are free parameters, with $i,j=1,2,3$. Furthermore, $Y_{\bold{r}}^{2k}$ are modular forms with $\bold{r}=\bold{1},\bold{1'},\bold{1''}$ representations of $A_{4}$ symmetry and $2k$ are modular weights.
We consider the limit $m_S \ll \Lambda_c$.
In this case, light neutrino masses are given by the double seesaw formula,
\begin{equation}
M_{\nu}= Y_{D}(M_{D}^{'}M_{S}^{-1}M_{D}^{'T})^{-1}Y_{D}^{T} \,.
\end{equation}

Note that non-renormalisable superpotential terms as,
\begin{equation}
\Phi \tilde{\Phi} Y_{\bold{1''}}^{(8+2k)} \tilde{\xi}^{12+2k}
\end{equation}
for $2k=2,4,...$, are suppressed due to the large power of $\tilde{\xi}$. {This could also be forbidden by introducing additional $Z_2$ as discussed in the end of section~3.2.}




\begin{thebibliography}{99}
\bibitem{Feruglio:2017spp}
F.~Feruglio,
[arXiv:1706.08749 [hep-ph]].



\bibitem{Dudas:1995eq}
E.~Dudas, S.~Pokorski and C.~A.~Savoy,
Phys. Lett. B \textbf{369} (1996), 255-261
[arXiv:hep-ph/9509410 [hep-ph]].

\bibitem{Leontaris:1997vw}
G.~K.~Leontaris and N.~D.~Tracas,
Phys. Lett. B \textbf{419} (1998), 206-210
[arXiv:hep-ph/9709510 [hep-ph]].

\bibitem{Dent:2001mn}
T.~Dent,
JHEP \textbf{12} (2001), 028
[arXiv:hep-th/0111024 [hep-th]].


\bibitem{Nilles:2020nnc}
H.~P.~Nilles, S.~Ramos-S\'anchez and P.~K.~S.~Vaudrevange,
JHEP \textbf{02} (2020), 045
doi:10.1007/JHEP02(2020)045
[arXiv:2001.01736 [hep-ph]].

\bibitem{Kobayashi:2020uaj}
T.~Kobayashi and H.~Otsuka,
Phys. Rev. D \textbf{102} (2020) no.2, 026004
doi:10.1103/PhysRevD.102.026004
[arXiv:2004.04518 [hep-th]].

\bibitem{Liu:2019khw}
X.~G.~Liu and G.~J.~Ding,
JHEP \textbf{08} (2019), 134
[arXiv:1907.01488 [hep-ph]].

\bibitem{Kobayashi:2018vbk}
T.~Kobayashi, K.~Tanaka and T.~H.~Tatsuishi,
Phys. Rev. D \textbf{98} (2018) no.1, 016004
[arXiv:1803.10391 [hep-ph]].

\bibitem{Kobayashi:2018wkl}
T.~Kobayashi, Y.~Shimizu, K.~Takagi, M.~Tanimoto, T.~H.~Tatsuishi and H.~Uchida,
Phys. Lett. B \textbf{794} (2019), 114-121
[arXiv:1812.11072 [hep-ph]].

\bibitem{Kobayashi:2019rzp}
T.~Kobayashi, Y.~Shimizu, K.~Takagi, M.~Tanimoto and T.~H.~Tatsuishi,
PTEP \textbf{2020} (2020) no.5, 053B05
[arXiv:1906.10341 [hep-ph]].

\bibitem{Okada:2019xqk}
H.~Okada and Y.~Orikasa,
Phys. Rev. D \textbf{100} (2019) no.11, 115037
[arXiv:1907.04716 [hep-ph]].

\bibitem{Criado:2018thu}
J.~C.~Criado and F.~Feruglio,
SciPost Phys. \textbf{5} (2018) no.5, 042
[arXiv:1807.01125 [hep-ph]].

\bibitem{Kobayashi:2018scp}
T.~Kobayashi, N.~Omoto, Y.~Shimizu, K.~Takagi, M.~Tanimoto and T.~H.~Tatsuishi,
JHEP \textbf{11} (2018), 196
[arXiv:1808.03012 [hep-ph]].

\bibitem{deAnda:2018ecu}
F.~J.~de Anda, S.~F.~King and E.~Perdomo,
Phys. Rev. D \textbf{101} (2020) no.1, 015028
[arXiv:1812.05620 [hep-ph]].

\bibitem{Okada:2018yrn}
H.~Okada and M.~Tanimoto,
Phys. Lett. B \textbf{791} (2019), 54-61
[arXiv:1812.09677 [hep-ph]].

\bibitem{Novichkov:2018yse}
P.~P.~Novichkov, S.~T.~Petcov and M.~Tanimoto,
Phys. Lett. B \textbf{793} (2019), 247-258
[arXiv:1812.11289 [hep-ph]].

\bibitem{Nomura:2019jxj}
T.~Nomura and H.~Okada,
Phys. Lett. B \textbf{797} (2019), 134799
[arXiv:1904.03937 [hep-ph]].

\bibitem{Okada:2019uoy}
H.~Okada and M.~Tanimoto,
Eur. Phys. J. C \textbf{81} (2021) no.1, 52
[arXiv:1905.13421 [hep-ph]].

\bibitem{Nomura:2019yft}
T.~Nomura and H.~Okada,
Nucl. Phys. B \textbf{966} (2021), 115372
[arXiv:1906.03927 [hep-ph]].

\bibitem{Ding:2019zxk}
G.~J.~Ding, S.~F.~King and X.~G.~Liu,
JHEP \textbf{09} (2019), 074
[arXiv:1907.11714 [hep-ph]].

\bibitem{Okada:2019mjf}
H.~Okada and Y.~Orikasa,
[arXiv:1907.13520 [hep-ph]].

\bibitem{Nomura:2019lnr}
T.~Nomura, H.~Okada and O.~Popov,
Phys. Lett. B \textbf{803} (2020), 135294
[arXiv:1908.07457 [hep-ph]].

\bibitem{Kobayashi:2019xvz}
T.~Kobayashi, Y.~Shimizu, K.~Takagi, M.~Tanimoto and T.~H.~Tatsuishi,
Phys. Rev. D \textbf{100} (2019) no.11, 115045
[erratum: Phys. Rev. D \textbf{101} (2020) no.3, 039904]
[arXiv:1909.05139 [hep-ph]].

\bibitem{Asaka:2019vev}
T.~Asaka, Y.~Heo, T.~H.~Tatsuishi and T.~Yoshida,
JHEP \textbf{01} (2020), 144
[arXiv:1909.06520 [hep-ph]].

\bibitem{Gui-JunDing:2019wap}
G.~J.~Ding, S.~F.~King, X.~G.~Liu and J.~N.~Lu,
JHEP \textbf{12} (2019), 030
[arXiv:1910.03460 [hep-ph]].

\bibitem{Zhang:2019ngf}
D.~Zhang,
Nucl. Phys. B \textbf{952} (2020), 114935
[arXiv:1910.07869 [hep-ph]].

\bibitem{Nomura:2019xsb}
T.~Nomura, H.~Okada and S.~Patra,
Nucl. Phys. B \textbf{967} (2021), 115395
[arXiv:1912.00379 [hep-ph]].

\bibitem{Wang:2019xbo}
X.~Wang,
Nucl. Phys. B \textbf{957} (2020), 115105
[arXiv:1912.13284 [hep-ph]].

\bibitem{Kobayashi:2019gtp}
T.~Kobayashi, T.~Nomura and T.~Shimomura,
Phys. Rev. D \textbf{102} (2020) no.3, 035019
[arXiv:1912.00637 [hep-ph]].

\bibitem{King:2020qaj}
S.~J.~D.~King and S.~F.~King,
JHEP \textbf{09} (2020), 043
[arXiv:2002.00969 [hep-ph]].

\bibitem{Ding:2020yen}
G.~J.~Ding and F.~Feruglio,
JHEP \textbf{06} (2020), 134
[arXiv:2003.13448 [hep-ph]].

\bibitem{Okada:2020rjb}
H.~Okada and M.~Tanimoto,
[arXiv:2005.00775 [hep-ph]].

\bibitem{Nomura:2020opk}
T.~Nomura and H.~Okada,
[arXiv:2007.04801 [hep-ph]].

\bibitem{Asaka:2020tmo}
T.~Asaka, Y.~Heo and T.~Yoshida,
Phys. Lett. B \textbf{811} (2020), 135956
[arXiv:2009.12120 [hep-ph]].

\bibitem{Okada:2020brs}
H.~Okada and M.~Tanimoto,
JHEP \textbf{03} (2021), 010
[arXiv:2012.01688 [hep-ph]].

\bibitem{Yao:2020qyy}
C.~Y.~Yao, J.~N.~Lu and G.~J.~Ding,
JHEP \textbf{05} (2021), 102
[arXiv:2012.13390 [hep-ph]].

\bibitem{Feruglio:2021dte}
F.~Feruglio, V.~Gherardi, A.~Romanino and A.~Titov,
JHEP \textbf{05} (2021), 242
[arXiv:2101.08718 [hep-ph]].

\bibitem{Penedo:2018nmg}
J.~T.~Penedo and S.~T.~Petcov,
Nucl. Phys. B \textbf{939} (2019), 292-307
[arXiv:1806.11040 [hep-ph]].

\bibitem{Novichkov:2018ovf}
P.~P.~Novichkov, J.~T.~Penedo, S.~T.~Petcov and A.~V.~Titov,
JHEP \textbf{04} (2019), 005
[arXiv:1811.04933 [hep-ph]].

\bibitem{deMedeirosVarzielas:2019cyj}
I.~de Medeiros Varzielas, S.~F.~King and Y.~L.~Zhou,
Phys. Rev. D \textbf{101} (2020) no.5, 055033
[arXiv:1906.02208 [hep-ph]].

\bibitem{Kobayashi:2019mna}
T.~Kobayashi, Y.~Shimizu, K.~Takagi, M.~Tanimoto and T.~H.~Tatsuishi,
JHEP \textbf{02} (2020), 097
[arXiv:1907.09141 [hep-ph]].

\bibitem{Criado:2019tzk}
J.~C.~Criado, F.~Feruglio and S.~J.~D.~King,
JHEP \textbf{02} (2020), 001
[arXiv:1908.11867 [hep-ph]].

\bibitem{King:2019vhv}
S.~F.~King and Y.~L.~Zhou,
Phys. Rev. D \textbf{101} (2020) no.1, 015001
[arXiv:1908.02770 [hep-ph]].

\bibitem{Wang:2019ovr}
X.~Wang and S.~Zhou,
JHEP \textbf{05} (2020), 017
[arXiv:1910.09473 [hep-ph]].

\bibitem{Wang:2020dbp}
X.~Wang,
Nucl. Phys. B \textbf{962} (2021), 115247
[arXiv:2007.05913 [hep-ph]].

\bibitem{Qu:2021jdy}
B.~Y.~Qu, X.~G.~Liu, P.~T.~Chen and G.~J.~Ding,
[arXiv:2106.11659 [hep-ph]].

\bibitem{Novichkov:2018nkm}
P.~P.~Novichkov, J.~T.~Penedo, S.~T.~Petcov and A.~V.~Titov,
JHEP \textbf{04} (2019), 174
[arXiv:1812.02158 [hep-ph]].

\bibitem{Ding:2019xna}
G.~J.~Ding, S.~F.~King and X.~G.~Liu,
Phys. Rev. D \textbf{100} (2019) no.11, 115005
[arXiv:1903.12588 [hep-ph]].

\bibitem{Ding:2020msi}
G.~J.~Ding, S.~F.~King, C.~C.~Li and Y.~L.~Zhou,
JHEP \textbf{08} (2020), 164
[arXiv:2004.12662 [hep-ph]].

\bibitem{Ding:2020zxw}
G.~J.~Ding, F.~Feruglio and X.~G.~Liu,
JHEP \textbf{01} (2021), 037
[arXiv:2010.07952 [hep-th]].

\bibitem{Froggatt:1978nt}
C.~D.~Froggatt and H.~B.~Nielsen,
Nucl. Phys. B \textbf{147} (1979), 277-298

\bibitem{Georgi:1974sy}
H.~Georgi and S.~L.~Glashow,
Phys. Rev. Lett. \textbf{32} (1974), 438-441

\bibitem{King:2017guk}
S.~F.~King,
Prog. Part. Nucl. Phys. \textbf{94} (2017), 217-256
[arXiv:1701.04413 [hep-ph]].

\bibitem{Ma:2001dn}
E.~Ma and G.~Rajasekaran,
Phys. Rev. D \textbf{64} (2001), 113012
[arXiv:hep-ph/0106291 [hep-ph]].

\bibitem{Bjorkeroth:2015ora}
F.~Bj\"orkeroth, F.~J.~de Anda, I.~de Medeiros Varzielas and S.~F.~King,
JHEP \textbf{06} (2015), 141
[arXiv:1503.03306 [hep-ph]].

\bibitem{Chen:2021zty}
P.~Chen, G.~J.~Ding and S.~F.~King,
JHEP \textbf{04} (2021), 239
[arXiv:2101.12724 [hep-ph]].

\bibitem{Du:2020ylx}
X.~Du and F.~Wang,
JHEP \textbf{02} (2021), 221
[arXiv:2012.01397 [hep-ph]].

\bibitem{Zhao:2021jxg}
Y.~Zhao and H.~H.~Zhang,
JHEP \textbf{03} (2021), 002
[arXiv:2101.02266 [hep-ph]].

\bibitem{King:2021fhl}
S.~F.~King and Y.~L.~Zhou,
JHEP \textbf{04} (2021), 291
[arXiv:2103.02633 [hep-ph]].

\bibitem{Ding:2021zbg}
G.~J.~Ding, S.~F.~King and C.~Y.~Yao,
[arXiv:2103.16311 [hep-ph]].

\bibitem{Ding:2021eva}
G.~J.~Ding, S.~F.~King and J.~N.~Lu,
[arXiv:2108.09655 [hep-ph]].

\bibitem{DeRujula:1980qc}
A.~De Rujula, H.~Georgi and S.~L.~Glashow,
Phys. Rev. Lett. \textbf{45} (1980), 413

\bibitem{Georgi:1980pw}
H.~Georgi, S.~L.~Glashow and M.~Machacek,
Phys. Rev. D \textbf{23} (1981), 783

\bibitem{Barr:1981qv}
S.~M.~Barr,
Phys. Lett. B \textbf{112} (1982), 219-222

\bibitem{Derendinger:1983aj}
J.~P.~Derendinger, J.~E.~Kim and D.~V.~Nanopoulos,
Phys. Lett. B \textbf{139} (1984), 170-176

\bibitem{Antoniadis:1987dx}
I.~Antoniadis, J.~R.~Ellis, J.~S.~Hagelin and D.~V.~Nanopoulos,
Phys. Lett. B \textbf{194} (1987), 231-235

\bibitem{Pati:1974yy}
J.~C.~Pati and A.~Salam,
Phys. Rev. D \textbf{10} (1974), 275-289
[erratum: Phys. Rev. D \textbf{11} (1975), 703-703]

\bibitem{Antoniadis:1988cm}
I.~Antoniadis and G.~K.~Leontaris,
Phys. Lett. B \textbf{216} (1989), 333-335

\bibitem{Kyae:2005nv}
B.~Kyae and Q.~Shafi,
Phys. Lett. B \textbf{635} (2006), 247-252
[arXiv:hep-ph/0510105 [hep-ph]].

\bibitem{Georgi:1979df}
H.~Georgi and C.~Jarlskog,
Phys. Lett. B \textbf{86} (1979), 297-300

\bibitem{deMedeirosVarzielas:2020kji}
I.~de Medeiros Varzielas, M.~Levy and Y.~L.~Zhou,
JHEP \textbf{11} (2020), 085
[arXiv:2008.05329 [hep-ph]].

\bibitem{IbanezUranga} Luis E. Ibanez and Angel M. Uranga, 
{\it `String Theory and Particle Physics
An Introduction to String Phenomenology'}, Cambridge University Press, 2012

\bibitem{Antusch:2013jca}
S.~Antusch and V.~Maurer,
JHEP \textbf{11} (2013), 115
[arXiv:1306.6879 [hep-ph]].

\bibitem{Esteban:2020cvm}
I.~Esteban, M.~C.~Gonzalez-Garcia, M.~Maltoni, T.~Schwetz and A.~Zhou,
JHEP \textbf{09} (2020), 178
[arXiv:2007.14792 [hep-ph]].

	\bibitem{nufit5}
	NuFIT 5.0 (2020), www.nu-fit.org.

\bibitem{Zhou:2014sya}
Y.~L.~Zhou,
[arXiv:1409.8600 [hep-ph]].

\bibitem{Zhou:2012ds}
Y.~L.~Zhou,
Phys. Rev. D \textbf{86} (2012), 093011
[arXiv:1205.2303 [hep-ph]].

\end{thebibliography}
\end{document}